\documentclass[12pt,letterpaper]{article}
\pdfoutput=1
\usepackage{graphicx}
\usepackage{subfig}
\usepackage{amsfonts,amssymb}

\newcommand{\be}{\begin{equation}}
\newcommand{\ee}{\end{equation}}
\newcommand{\ben}{\begin{displaymath}}
\newcommand{\een}{\end{displaymath}}
\newcommand{\bea}{\begin{eqnarray}}
\newcommand{\eea}{\end{eqnarray}}
\newcommand{\bean}{\begin{eqnarray*}}
\newcommand{\eean}{\end{eqnarray*}}

\newcommand{\ads}[1]{\mbox{${AdS}_{#1}$}}

\newcommand{\eg}{{\it e.g.}}
\newcommand{\ie}{{\it i.e.}}

\newcommand{\commentout}[1]{}

\newcommand{\beq}{\begin{equation}}
\newcommand{\eeq}{\end{equation}}
\newcommand{\beqr}{\begin{displaymath}}
\newcommand{\eeqr}{\end{displaymath}}
\newcommand{\beqa}{\begin{eqnarray}}
\newcommand{\eeqa}{\end{eqnarray}}
\newcommand{\beqar}{\begin{eqnarray*}}
\newcommand{\eeqar}{\end{eqnarray*}}
\newcommand{\cN}{{\cal N}}

\newcommand{\cO}{{\cal O}}

\newcommand{\half}{\ensuremath{\frac{1}{2}}}

\newcommand{\N}[1]{\ensuremath{\cN=#1}}

\newcommand{\cU}{\ensuremath{\mathcal{U}}}

\begin{document}

\title{\LARGE \bf Notes on Euclidean Wilson loops and Riemann Theta functions}

\author{Riei Ishizeki, Martin Kruczenski, Sannah Ziama\thanks{E-mail: \texttt{rishizek@purdue.edu, markru@purdue.edu, sziama@purdue.edu}}\\
        Department of Physics, Purdue University,  \\
        525 Northwestern Avenue, W. Lafayette, IN 47907-2036. }

\maketitle

\begin{abstract}

The AdS/CFT correspondence relates Wilson loops in \N{4} SYM theory to minimal 
area surfaces in \ads{5} space. In this paper we consider the case of Euclidean flat Wilson loops which 
are related to minimal area surfaces in Euclidean \ads{3} space. Using known mathematical results for such
minimal area surfaces we describe an infinite parameter family of analytic solutions for closed Wilson loops. 
The solutions are given in terms of Riemann theta functions and the validity of the equations of motion is proven
based on the trisecant identity. The world-sheet has the topology of a disk and the renormalized area is written
as a finite, one-dimensional contour integral over the world-sheet boundary. An example is discussed in detail 
with plots of the corresponding surfaces. Further, for each Wilson loops we explicitly construct a one parameter family of 
deformations that preserve the area. The parameter is the so called spectral parameter.  Finally, for genus three 
we find a map between these Wilson loops and closed curves inside the Riemann surface.

\end{abstract}

\clearpage
\newpage



\section{Introduction}
\label{intro}

 One of the first results of the AdS/CFT correspondence \cite{malda} was the computation of Wilson loops and from there the quark anti-quark potential
 as done by Maldacena, Rey and Yee \cite{MRY}. Although much work was devoted to the computation of Wilson loops only few explicit examples are 
 known of minimal area surfaces in \ads{5} space. In the case of closed Euclidean Wilson loops (with constant scalar) the most studied one is the  
circular Wilson loop \cite{cWL} which is dual to a half-sphere. The only other one we are aware of is the two intersecting arcs (lens shaped) \cite{lens}.
For infinite Wilson loops, parallel lines \cite{MRY} and the cusp are known \cite{DGO}. In the case of multiple contours as for example two concentric 
circles, interesting results were found using integrability \cite{DF}. In the language we use here they correspond to elliptic functions which appear 
for genus one $g=1$.
 In the case of Minkowski signature AdS space and in particular light-like lines more is known starting with the light-like cusp \cite{cusp} and 
culminating  with a large recent activity \cite{scatampl} in relation to scattering amplitudes following \cite{AM, AM2}.

  In this paper we point out that in the case of flat  Euclidean Wilson loops which are dual to minimal area surfaces in Euclidean \ads{3}, much 
can be done  by using known results from the mathematical literature \cite{BB}. In fact, an infinite parameter family of solution is known in terms 
of Riemann theta functions. 
  This type of construction using theta functions is described in detail in \cite{BBook} and was already used in the case of strings moving 
in $t\times S^3$  by Dorey and Vicedo \cite{DV} and in the case of an Euclidean world-sheet inside \ads{3} space by Sakai and Satoh in \cite{SS}. 
Here we consider 
 Euclidean Wilson loops inside Euclidean \ads{3} and rederive the original results by perhaps more pedestrian methods based on the trisecant identity 
 for theta functions. In this way theta functions are thought as special functions whose properties fit well with the equations of motion of the string in \ads{3} 
 space much in the same way as trigonometric functions fit the harmonic oscillator equation. Each theta function and therefore each Wilson loop is associated 
 with an auxiliary Riemann surface of given genus $g$. A relatively simple formula is derived for the area and an example for genus $g=3$ is worked out in detail. Perhaps our main contribution is to find closed Wilson loops and to derive a formula for the renormalized area that follows the AdS/CFT prescription. 
  The calculations are done at the classical level, it should be interesting to extend them for example to one-loop as can be done in the case of the circular 
Wilson loop \cite{olcWL}.
 
  This paper is organized as follows. We start by writing the equations of motion and use the Pohlmeyer reduction procedure to simplify the equations 
and arrive at the cosh-Gordon equation plus a set of linear equations. In the following section we review the properties of the theta functions and show how 
they can be used to solve the equations of motion and compute the regularized area.  Finally we construct a particular example of genus three where we 
show that there are closed Wilson loops that can be described by this method. We plot the corresponding surfaces and compute the areas. Besides, we also describe 
the mapping of the Wilson loop into a curve embedded inside the Riemann surface. In the last section we give our conclusions. 

\section{Equations of motion}

In this section we write the equations of motion and simplify them using the Pohlmeyer reduction \cite{Pohlmeyer}. In the context of Minkowski space-time
this procedure was used by Jevicki and Jin \cite{JJ} to find new spiky string \cite{spiky} solutions and by Alday and Maldacena \cite{AM2} to compute certain light-like
Wilson loops.   In the case of Euclidean \ads{3} that we are interested in here, we can use embedding coordinates $X_{\mu=0\ldots 3}$ parameterizing a 
space $R^{3,1}$ and subjected to the constraint
\beq
  X_0^2 - X_1^2 - X_2^2 - X_3^2 = 1\ ,
\label{Xsq1}
\eeq
with an obvious  $SO(3,1)\equiv SL(2,\mathbb{C})$ global invariance. The space has an $S^2$ boundary at infinity.
Other useful coordinates are Poincare coordinates $(X,Y,Z)$ given by:
\beq
 X+iY= \frac{X_1+iX_2}{X_0-X_3}, \ \ \ \ Z= \frac{1}{X_0-X_3}.
\eeq
The boundary is now an $R^2$ space and located at $Z=0$.
A string is parameterized by world-sheet coordinates $\sigma_a=(\sigma,\tau)$ or equivalently complex coordinates
$z=\sigma+i\tau$, $\bar{z}=\sigma-i\tau$. The action in conformal gauge is given by
\beqa
 S &=& \half \int \left( \partial X_\mu \bar{\partial} X^\mu - \Lambda (X_\mu X^\mu -1)\right) \ d\sigma\, d\tau\\
    &=& \half \int \frac{1}{Z^2}\left(\partial_a X \partial^a X +\partial_a Y \partial^a Y +\partial_a Z \partial^a Z \right) \ d\sigma\, d\tau,
   \label{action}
\eeqa
where $\Lambda$ is a Lagrange multiplier and the $\mu$ indices are raised and lowered with the $R^{3,1}$ metric.
An Euclidean classical string is given by functions $X_\mu(z,\bar{z})$ obeying the equations of motion:
\beq
  \partial \bar{\partial} X_\mu = \Lambda X_\mu\ ,
\label{eomX}
\eeq
where $\Lambda$, the Lagrange multiplier is given by 
\beq
\Lambda = -\partial X_\mu \bar{\partial} X^{\mu}. \label{Lcomp}
\eeq
These equations  should be supplemented by the Virasoro constraints which read
\beq
 \partial X_\mu \partial X^\mu = 0 = \bar{\partial} X_\mu \bar{\partial} X^\mu.
\label{Vc1}
\eeq
Later on we will be interested in finding the solutions in Poincare coordinates $(X,Y,Z)$ but for the moment it is convenient
to study the problem in embedding coordinates $X_\mu$. We can rewrite the equations using the matrix
\beq
\mathbb{X} = \left(\begin{array}{cc} X_0+X_3 & X_1-i X_2 \\ X_1 + i X_2 & X_0-X_3 \end{array} \right) = X_0 + X_i \sigma^i \ ,
\eeq
where $\sigma^i$ denote the Pauli matrices. Notice also that Poincare coordinates are simply given by
\beq
Z = \frac{1}{\mathbb{X}_{22}} , \ \ \ X+iY =\frac{\mathbb{X}_{21}}{\mathbb{X}_{22}}.
\label{PoinX}
\eeq
The matrix $\mathbb{X}$ satisfies
\beq
\mathbb{X}^\dagger = \mathbb{X} , \ \ \det \mathbb{X} = 1, \ \ \ \partial\bar{\partial} \mathbb{X}=\Lambda \mathbb{X}, \ \ 
\det(\partial\mathbb{X})=0=\det(\bar{\partial}\mathbb{X})\ ,
\eeq
as follows from the definition of $\mathbb{X}$, the constraint (\ref{Xsq1}), the equations of motion (\ref{eomX}) and the Virasoro constraints (\ref{Vc1}). 
We can solve the constraint $\mathbb{X}^\dagger=\mathbb{X}$ by writing 
\beq
\mathbb{X}=\mathbb{A}\mathbb{A}^\dagger, \ \ \ \ \det \mathbb{A}=1 , \ \ \ \ \mathbb{A} \in SL(2,\mathbb{C}).
\eeq
The equations of motion have a global $SL(2,\mathbb{C})\equiv SO(3,1)$ symmetry under which
\beq
\mathbb{X} \rightarrow U\mathbb{X}U^\dagger, \ \ \ \mathbb{A}\rightarrow U\mathbb{A}, \ \ \ U\in SL(2,\mathbb{C}).
\eeq
In the new variable there is an $SU(2)$ gauge symmetry
\beq
 \mathbb{A} \rightarrow \mathbb{A} \cU, \ \ \ \cU(z,\bar{z})\in SU(2)\ ,
\eeq
since this leaves $\mathbb{X}$ invariant. We can define the current
\beq
J = \mathbb{A}^{-1} \partial \mathbb{A}, \ \ \ \bar{J}=\mathbb{A}^{-1} \bar{\partial} \mathbb{A}\ ,
\label{Jdef}
\eeq
which is invariant under the global symmetry and, under the local symmetry transform as
\beq
J \rightarrow \cU^{\dagger} J\cU + \cU^\dagger \partial \cU, \ \ \ \bar{J} \rightarrow \cU^{\dagger} \bar{J} \cU + \cU^\dagger \bar{\partial} \cU.
\eeq
From the definition of $J$, $\bar{J}$, the property that $\det \mathbb{A}=1$, the equations of motion and the constraints we find:
\beqa
 \bar{\partial} J -\partial\bar{J} + [\bar{J},J] &=& 0\ ,\\
 \mbox{Tr} J = \mbox{Tr} \bar{J} &=& 0\ ,\\
 \partial ( \bar{J} + J^\dagger ) + \half [J-\bar{J}^\dagger,\bar{J}+J^\dagger] &=& 0\ ,\\
 \det (\bar{J}+J^\dagger) &=& 0\ ,
\eeqa
and the corresponding equations found by hermitian conjugations of the ones given. In the third equation we used that for $SU(2)$ currents we
have for example:
\beq
J \bar{J} = \half [J,\bar{J}] + \half \mbox{Tr}\left( J\bar{J}\right)\ ,
\eeq
and similarly for the other products. The trace part gives the Lagrange multiplier $\Lambda$ as:
\beq
\Lambda = \half \mbox{Tr} \left(\left(J+\bar{J}^\dagger\right) \left(J^\dagger+\bar{J}\right)\right)  \label{LJ} \ ,
\eeq
which does not provide an equation but is useful later to determine the world-sheet metric.
 From the form of the equations it seems convenient to define 
\beq
 \mathcal{A} =\half(\bar{J}+J^\dagger), \ \ \ \mathcal{B}=\half(J-\bar{J}^\dagger).
 \label{ABdef}
\eeq
The equations read now
\beqa
 \mbox{Tr} \mathcal{A} = \mbox{Tr} \mathcal{B} &=& 0\ , \label{eqAB1}\\
 \det \mathcal{A} &=&0 \ ,\label{eqAB2} \\
\partial \mathcal{A} + [\mathcal{B},\mathcal{A}] &=& 0 \ , \label{eqAB3}\\
\bar{\partial} \mathcal{B} + \partial \mathcal{B}^\dagger &=& [\mathcal{B}^\dagger,\mathcal{B}] + [\mathcal{A}^\dagger,\mathcal{A}]. \label{eqAB4}
\eeqa
The $SU(2)$ gauge symmetry acts on these currents as
\beq
 \mathcal{A} \rightarrow \cU^\dagger \mathcal{A} \cU, \ \ \ \  \mathcal{B} \rightarrow \cU^\dagger \mathcal{B} \cU  +\cU^\dagger\partial \cU,\ \ \ 
 \cU(z,\bar{z}) \in SU(2).
\eeq
In a sense, $\mathcal{B}$ plays the role of a gauge field.
Since $ \mbox{Tr} \mathcal{A} = 0 $ we can write in terms of Pauli matrices $\sigma^j$:
\beq
\mathcal{A} = \left(\mathcal{A}^{(1)}_j + i \mathcal{A}^{(2)}_j\right) \sigma^j \ ,
\eeq
with $\mathcal{A}^{1,2}$ two real three-dimensional vectors. The property $\det \mathcal{A} =0$ implies that they are orthogonal and of the same 
length. The $SU(2)$ gauge symmetry acts on them as three-dimensional rotation so we can take $\mathcal{A}^{(1)}$ to be along the $\hat{x}$ axis, and
$\mathcal{A}^{(2)}$ along the $\hat{y}$ axis. In that way we can choose the gauge such that
\beq
\mathcal{A} = \half e^{\alpha(z,\bar{z})} (\sigma_1+ i \sigma_2) = e^{\alpha(z,\bar{z})} \sigma_+ \ ,
\eeq
where $\alpha(z,\bar{z})$ is a real function and $\sigma_+ = \half(\sigma_1+i\sigma_2) = \left(\begin{array}{cc} 0&1\\ 0 & 0   \end{array}\right)$.
Equation $\partial \mathcal{A} + [\mathcal{B},\mathcal{A}] =0$ uniquely implies that
\beq
\mathcal{B} = -\half \partial \alpha \sigma_z + b_2(z,\bar{z}) \sigma_+ \ ,
\eeq
for some function $b_2(z,\bar{z})$.
Finally the equation for $\mathcal{B}$ implies that $b_2=f(z) e^{-\alpha}$ for an arbitrary holomorphic function $f(z)$. It also implies that
\beq
\partial\bar{\partial}\alpha = e^{2\alpha} +f \bar{f} e^{-2\alpha} .
\label{alphafeq}
\eeq
So, up to a gauge transformation the most general solution is given by
\beqa
 \mathcal{A} &=& e^{\alpha} \sigma_+ \ ,\\
 \mathcal{B} &=& -\half \partial \alpha \sigma_z + f(z) e^{-\alpha} \sigma_+ \ ,
\eeqa
with $\alpha$ satisfying eq.(\ref{alphafeq}).
Finally the function $f(z)$ can be locally eliminated by changing to world-sheet coordinates $w$ such that $dw=\sqrt{f} dz$. If we further redefine
$\alpha\rightarrow \alpha + \frac{1}{4}\ln(f\bar{f})$ and do a  gauge transformation with $\cU=e^{\frac{i\phi}{2}\sigma_z}$ 
(where $i\phi=\frac{1}{4}\, f/ \bar{f}$) then the result is equivalent to setting $f=1$. 

From the equations of motion for $\mathcal{A}$ and $\mathcal{B}$, namely eqns.(\ref{eqAB1})-(\ref{eqAB4}) it can be seen that they are invariant 
under multiplying $\mathcal{A}$ by a constant of modulus one that we call $\bar{\lambda}$ ($|\lambda|=1$). We get:
\beqa
 \mathcal{A} &=& \bar{\lambda} e^{\alpha} \sigma_+ \ ,\\
 \mathcal{B} &=& -\half \partial \alpha \sigma_z + e^{-\alpha} \sigma_+ \ ,\\
 \partial\bar{\partial}\alpha &=& 2 \cosh(2\alpha) \ ,
\eeqa
where we set $f=1$ by the reasons indicated before. 
The constant $\lambda$ can be eliminated by a gauge transformation and a redefinition of $\alpha$ but we keep it for later convenience. It is called
the spectral parameter and should not be confused with the coupling constant in the dual gauge theory.  

Having computed $\mathcal{A}$ and $\mathcal{B}$ we can use eq.(\ref{ABdef}) to reconstruct $J$ and $\bar{J}$ obtaining:
\beq
J=\left(\begin{array}{cc}-\half\partial\alpha & e^{-\alpha}\\ \lambda e^{\alpha} & \half \partial\alpha  \end{array}\right), \ \ \ \ 
\bar{J}=\left(\begin{array}{cc}\half \bar{\partial}\alpha & \bar{\lambda} e^{\alpha}\\ -e^{-\alpha} & -\half \bar{\partial}\alpha  \end{array}\right) .\label{JJbar}
\eeq
Finally we should use eq.(\ref{Jdef}) to compute $\mathbb{A}$. Summarizing we need first to solve the equation:
\beq
  \partial\bar{\partial}\alpha = 2\cosh 2\alpha\ , \label{alphaeqn} 
\eeq
then plug $\alpha$ into the definitions for $J$, $\bar{J}$, namely eq.(\ref{JJbar}), and solve for $\mathbb{A}$:
\beqa
  \partial \mathbb{A} &=& \mathbb{A} J   \ ,    \label{A1eqn} \\
  \bar{\partial} \mathbb{A} &=& \mathbb{A} \bar{J} .  \label{A2eqn}
 \eeqa
  Finally, the string solution is determined as $\mathbb{X}=\mathbb{A}\mathbb{A}^\dagger$.  The equation for $\alpha$ is non-linear but the ones for $\mathbb{A}$  are linear since $J$, $\bar{J}$ are known once $\alpha$ is known. This is the main idea of the Pohlmeyer reduction \cite{Pohlmeyer} which we rederive here as it applies to our  particular problem. Similar considerations in the context of string theory are well-known, for example see \cite{JJ}, \cite{DDJK}, \cite{AM2}, \cite{SS}.
  
 Notice that, since $\mbox{Tr}J=\mbox{Tr}\bar{J}=0$ we 
 automatically find that $\det \mathbb{A}$ is constant independent of  $z,\bar{z}$. However we need $\det \mathbb{A}=1$ so we just need to normalize 
 $\mathbb{A}$ dividing by an appropriate constant. Furthermore it is convenient to write
 \beq
 \mathbb{A} = \left(\begin{array}{cc}\psi_1 & \psi_2\\ \tilde{\psi}_1 & \tilde{\psi}_2\end{array}\right)\ ,
 \eeq
where the vectors $\psi=(\psi_1,\psi_2)$ and $\tilde{\psi}=( \tilde{\psi}_1, \tilde{\psi}_2)$  are linearly independent and satisfy
\beq
\partial \psi = \psi J, \ \ \ \bar{\partial} \psi = \psi \bar{J}\ , \label{p12eq}
\eeq
and the same for $\tilde{\psi}$. They have to be linearly independent so the determinant $(\psi_1 \tilde{\psi}_2 -\psi_2 \tilde{\psi}_1)$ is non vanishing 
(but is constant as discussed before). Even with these conditions there is a certain ambiguity in choosing $\psi$, $\tilde{\psi}$ but those boil down to
$SL(2,\mathbb{C})\equiv SO(3,1)$ transformations of $\mathbb{X}$.

\section{Solutions}

As shown in \cite{BB} solutions to eqns.(\ref{alphaeqn},\ref{A1eqn},\ref{A2eqn}) can be found using theta functions. In this section we rederive 
the results of \cite{BB} using a perhaps more pedestrian way. The method we use is to consider the theta functions as special functions whose derivatives are such that
they are good candidates to solve eqns.(\ref{alphaeqn},\ref{A1eqn},\ref{A2eqn}) in much the same way as trigonometric functions are good candidates to solve the harmonic oscillator.
The equations are solved by simple substitution and adjustment of the parameters. To start we need to review some properties of the theta functions.

\subsection{Riemann theta functions and their properties.}

There is a vast literature on Riemann theta functions \cite{ThF}. In this section we review the minimal knowledge necessary to find 
solutions to the equations. We follow the notation of \cite{FK} which also gives a good introduction to Riemann surfaces.   Notice also
that in the next section we develop an example in detail which can be read in parallel with this section.
 Consider a compact Riemann surface of genus $g$ with fundamental cycles $a_i$, $b_i$ ($i=1\ldots g$) and intersections
\beq
a_i \circ a_j =0 = b_i \circ b_j, \ \ \ a_i\circ b_j = \delta_{ij}\ ,
\eeq
which means that $a_i$ only intersects $b_i$. The Riemann surface is taken to be a hyperelliptic one defined by the function:
\beq
 \mu(\lambda) = \sqrt{\lambda \prod_{j=1}^{2g} (\lambda-\lambda_j)} .
\eeq
The square root has cuts with branching points at $0,\infty, \lambda_j$ but is well defined in a double cover of the complex plane. 
This double cover is the Riemann surface we consider.  For all values of $\lambda\neq 0,\infty, \lambda_j$  there are two points on the Riemann surface,
one in the upper sheet, and one in the lower sheet.  

Consider now $\omega_{i=1\ldots g}$ to be the unique basis of holomorphic abelian differentials 
satisfying $\oint_{a_i} \omega_j = \delta_{ij}$, and define the $g\times g$ period matrix
\beq
 \Pi_{ij} = \oint_{b_i} \omega_j.
 \label{Pidef}
\eeq
It can be proved that $\Pi$ is a symmetric matrix and its imaginary part is positive definite allowing the definition of an associated $\theta$-function
\beq
\theta(\zeta) = \sum_{n\in \mathbb{Z}^g} e^{ 2\pi i \left(n^t\Pi n+n^t \zeta\right)}.
\eeq
The arguments of the $\theta$-function are $\zeta$ which is a vector in $\mathbb{C}^g$ and the period matrix $\Pi$ (which we consider fixed and therefore do not
explicitly write as an argument).  The sum is done over all $n\in\mathbb{Z}^g$, that is all order $g$ vectors with integer components. All vectors (\eg\ $n,\zeta$) are 
taken to be column vectors (and therefore their transposes $n^t,\zeta^t$ are row vectors). 
 Simple but important properties of the theta function are
\beq
\theta(-\zeta)=\theta(\zeta)\ ,
\eeq 
and the (quasi)-periodicity:
\beq
\theta\left(\zeta+\Delta_{2}+\Pi \Delta_{1}\right) =e^{-2\pi i\left[\Delta_{1}^t \zeta+\half\Delta_{1}^t\Pi \Delta_{1}\right]} 
 \theta(\zeta)\ ,
\eeq
where $\Delta_1,\Delta_2\in \mathbb{Z}^g$, namely are vectors with integer components.
To shorten some equations, it is also useful to define the $\theta$ function with characteristics:
\beq
 \hat{\theta}(\zeta)=\theta\left[\begin{array}{c} \Delta_1 \\ \Delta_2 \end{array}\right](\zeta)=
\exp\left\{2\pi i\left[\frac{1}{8}\Delta_1^t \Pi \Delta_1+\half \Delta_1^t \zeta +\frac{1}{4}\Delta_1^t \Delta_2 \right]\right\} 
          \theta\left(\zeta+\half\Delta_2+\half\Pi \Delta_1\right)\ ,
\eeq
where again $\Delta_1,\Delta_2\in \mathbb{Z}^g$. 
 We introduced the notation $\hat{\theta}$  for this function
because, in the rest of the paper, $\Delta_1$ and $\Delta_2$ are fixed vectors. In particular from now on we are going to consider that $\Delta_1^t \Delta_2$ is an
odd integer which is also described as saying that $\left[\begin{array}{c} \Delta_1 \\ \Delta_2 \end{array}\right]$ is an odd characteristic.
In such case we have
\beq
\hat{\theta}(-\zeta) = e^{i\pi\Delta_1^t \Delta_2} \hat{\theta}(\zeta) = -\hat{\theta}(\zeta)\ ,
\eeq
as can be derived from the definition of $\hat{\theta}$ and we used that, in our case, $\Delta_1^t\Delta_2$ is odd. In particular this implies
\beq
\hat{\theta}(0)=0 \ \ \ \Rightarrow \ \ \ \theta\left(\half\Delta_2+\half\Pi \Delta_1\right)=0\ ,
\eeq
namely the vector $a=\half\Delta_2+\half\Pi \Delta_1$ is a zero of the theta function.
  The (quasi)-periodicity of the theta function implies that
\beq
\theta\left[\begin{array}{c} \Delta_1 +2 \varepsilon_1 \\ \Delta_2 +2 \varepsilon_2 \end{array}\right](\zeta) = 
e^{i\pi\Delta_1^t \varepsilon_2} \theta\left[\begin{array}{c} \Delta_1 \\ \Delta_2 \end{array}\right](\zeta)\ ,
\eeq 
for any $\varepsilon_{1,2}\in \mathbb{Z}^g$. Therefore it only makes sense to consider $\Delta_{1,2}$ modulus two, namely its components being zero or one. 

The most important property of the theta functions that we need in this paper is Fay's trisecant identity:
\begin{equation}
 \label{eq:fay}
\theta(\zeta)\; \theta\left(\zeta+\int_{p_{2}}^{p_1}\!\!\!\! \omega+\int_{p_3}^{p_{4}}\!\!\!\!\omega\right) =
\gamma_{1234}\, \theta\Big(\zeta+\int_{p_{2}}^{p_{1}}\!\!\!\!\omega \Big)\, \theta\Big(\zeta+\int_{p_{3}}^{p_{4}}\!\!\!\!\omega \Big)
 +\gamma_{1324}\, \theta\Big(\zeta+\int_{p_{3}}^{p_{1}}\!\!\!\!\omega \Big) \,\theta\Big(\zeta+\int_{p_{2}}^{p_{4}}\!\!\!\!\omega \Big) \ ,
\end{equation}
with
\beq
 \gamma_{ijkl}=\frac{\theta(a+\int_{p_{k}}^{p_{i}}\!\!\omega)\, \theta(a+\int_{p_{l}}^{p_{j}}\!\!\omega)}
      {\theta(a+\int_{p_{l}}^{p_{i}}\!\!\omega)\, \theta(a+\int_{p_{k}}^{p_{j}}\!\!\omega)} .
\eeq
In these formulas $p_j$ are points on the Riemann surface, and $a$ is a non-singular zero of the Riemann theta function, \ie\ the function is zero but not its gradient. 
In particular we are going to use $a=\half\Delta_2+\half\Pi \Delta_1$ which is a zero as noticed before. Also notice that the contour integrals
$\int_{p_a}^{p_b} \omega_j $ define a vector which from now on, following standard convention will be abbreviated as:
\beq
 \int_{p_a}^{p_b} \omega_j \rightarrow  \int_{p_a}^{p_b}.
 \eeq
The function $\gamma$ may be viewed as a generalization of the cross-ratio function on $\mathbb{CP}^{1}$ to functions on Riemann surfaces. 
Some immediate properties of these function are:
\beq
\gamma_{1233}\,=\,\gamma_{1134}\,=\,1, \ \   \gamma_{2134}\,=\,\gamma_{1234}^{-1},\ \ \  \gamma_{1214} =0=\gamma_{1232}.
\label{gammaid}
\eeq
One important use of the Fay's Trisecant formula is that it provides a direct way of obtaining directional derivatives of theta functions or of ratios 
of them. Taking the  derivative with respect to $p_{1}$ and then letting $p_{2}\rightarrow p_1$  we get 
\beqa
 \lefteqn{D_{p_{1}}\!\ln\left[\frac{\theta(\zeta)}{\theta(\zeta+\int_{p_{3}}^{p_{4}})}\right] = 
 - D_{p_{1}}\!\ln\Big[\frac{\theta(a+\int_{p_{3}}^{p_{1}})}{\theta(a+\int_{p_{4}}^{p_{1}})}\Big] } && \nonumber \\
&& \ \ \ \ \ \ \ \ \ \ \ \  -    \frac{D_{p_{1}}\!\theta(a)\:\theta\left(a+\int_{p_{4}}^{p_{3}}\right)} 
   {\theta\Big(a+\int_{p_{4}}^{p_{1}}\Big)\:\theta\Big(a+\int_{p_{1}}^{p_{3}}\Big)}
   \frac{\theta\Big(\zeta+\int_{p_{3}}^{p_{1}}\Big)\:\theta\Big(\zeta+\int_{p_{1}}^{p_{4}}\Big)}
    {\theta(\zeta)\:\theta\Big(\zeta+\int_{p_{3}}^{p_{4}}\Big)}.
\label{eq:finaldp1}
\eeqa
Here $D_{p_1}$ indicates a directional derivative defined as (summation over $j$ implied):
\beq
 D_{p_1} F(\zeta) = \omega_j(p_1) \frac{\partial F(\zeta)}{\partial \zeta_j} \ ,
\eeq
and should not be confused with a derivative with respect to $p_1$ that, if appears, we will denote as $\partial_{p_1}$. Also, the final expression is simplified
using the identities (\ref{gammaid}). 
We can further derive with respect to $p_3$ and take $p_4\rightarrow p_3$ obtaining:
\beq
D_{p_3p_1} \ln \theta(\zeta) = D_{p_3p_1} \ln \theta\left(a+\int_{p_{3}}^{p_{1}} \right)
                                 - \frac{ D_{p_{1}}\!\theta(a) D_{p_3}\theta\left(a\right)} 
   {\theta\Big(a+\int_{p_{3}}^{p_{1}}\Big)\:\theta\Big(a+\int_{p_{1}}^{p_{3}}\Big)}
   \frac{\theta\Big(\zeta+\int_{p_{3}}^{p_{1}}\Big)\:\theta\Big(\zeta+\int_{p_{1}}^{p_{3}}\Big)}
    {\theta^2(\zeta)}.
    \label{Thetapp}
\eeq
This summarizes the basic properties we need. Much more is known about these functions as can be found in the references \cite{FK}, \cite{ThF}.

\subsection{Solution to cosh-Gordon equation}

Eq.(\ref{Thetapp}) shows that the second derivative of the logarithm of a theta function contains the theta function. So solutions of 
\beq
  \partial\bar{\partial}\alpha = 2\cosh 2\alpha =  e^{2\alpha} + e^{-2\alpha}\ ,
\eeq
should naturally be sought as logs of theta functions. To eliminate the constant term in eq.(\ref{Thetapp}) we subtract two such derivatives and get
\beqa
\lefteqn{D_{p_1p_3} \ln \frac{\theta(\zeta)}{\theta(\zeta+\int_{p_1}^{p_3})} =} \ \ \ \ \ \  && \\ && 
 \frac{D_{p_1}\theta(a)D_{p_3}\theta(a)}{\theta(a+\int_{p_3}^{p_1})\theta(a+\int_{p_1}^{p_3})}
 \left\{\frac{\theta(\zeta+2\int_{p_1}^{p_3})\theta(\zeta)}{\theta^2(\zeta+\int_{p_1}^{p_3})}
 -\frac{\theta(\zeta+\int_{p_1}^{p_3}-2\int_{p_1}^{p_3})\theta(\zeta+\int_{p_1}^{p_3})}{\theta^2(\zeta)}\right\} . \nonumber
\eeqa
To get back the same theta functions we need to exploit their periodicity and therefore require
\beq
2\int_{p_1}^{p_3}\!\!\omega = \Delta_2 + \Pi \Delta_1\ ,
\eeq
where $\Delta_2, \Delta_1$ are integer vectors and $\Pi$ is the period matrix in eq.(\ref{Pidef}). This gives
\beq
\theta(\zeta+2\int_{p_1}^{p_3}) = e^{-2\pi i \Delta_1^t \zeta -i \pi \Delta_1^t \Pi \Delta_1} \theta(\zeta).
\eeq
We obtain
\beqa
\lefteqn{D_{p_1p_3} \ln \frac{\theta(\zeta)}{\theta(\zeta+\int_{p_1}^{p_3})} =
\frac{D_{p_1}\theta(a)D_{p_3}\theta(a)}{\theta(a+\int_{p_3}^{p_1})\theta(a+\int_{p_1}^{p_3})} e^{-\frac{i\pi}{2} \Delta_1^t \Pi \Delta_1}
} \ \ \ \ \ \   && \label{aeq1} \\ && 
\times \left\{e^{-2\pi i \Delta_1^t\zeta} e^{-\frac{i\pi}{2}\Delta_1^t\Pi\Delta_1}\frac{\theta^2(\zeta)}{\theta^2(\zeta+\int_{p_1}^{p_3})}
 -e^{i\pi\Delta_1^t\Delta_2}e^{2\pi i \Delta_1^t\zeta} e^{\frac{i\pi}{2}\Delta_1^t\Pi\Delta_1}\frac{\theta^2(\zeta+\int_{p_1}^{p_3})}{\theta^2(\zeta)}\right\} .\nonumber
\eeqa
We should now choose $p_1$, $p_3$ and the path of integration between them such that $\Delta_1^t \Delta_2$ is odd so that $e^{i\pi\Delta_1^t\Delta_2}=-1$. Then 
we take
\beq
e^{2\alpha} = - e^{-2\pi i \Delta_1^t\zeta - \frac{i\pi}{2} \Delta_1^t\Pi\Delta_1} \frac{\theta^2(\zeta)}{\theta^2(\zeta+\int_{p_1}^{p_3})} 
                    =\frac{\theta^2(\zeta)}{\hat{\theta}^2 (\zeta)}\ ,
\label{alphasol}
\eeq
and 
\beq
\zeta=2\omega(p_1) \bar{z} + 2 \omega(p_3) z.
\eeq
The last choice results in $\partial_z \zeta= 2 D_{p_3} \zeta$, $\bar{\partial}\zeta = 2 D_{p_1} \zeta$. 
The correct normalization for $\zeta$ and $\alpha$ follows from the result
\beq
\frac{D_{p_1}\theta(a)D_{p_3}\theta(a)}{\theta^2(0)} = -\frac{1}{4} e^{-i\frac{\pi}{2}\Delta_1^t\Pi\Delta_1}\ ,
\eeq
which is explained in the appendix. In any case it should be clear at this point that the overall normalization of $\alpha$  can always be adjusted so 
that eq.(\ref{aeq1}) becomes the cosh-Gordon eq. $\partial\bar{\partial}{\alpha} = 2\cosh 2\alpha$.
The final and very important point is that the theta functions are generically complex but $\alpha$ should be real. Again, following \cite{BB} we impose a 
reality condition as follows. Suppose there is a $g\times g$ symmetric matrix $T$ such that
\beq
\bar{\Pi} = -T \Pi T, \ \ \ \ \bar{\zeta} =-T\zeta, \ \ \ \ T^2=1.
\eeq
Then, it is easy to prove, using the definition of the theta function that $\theta(\zeta)$, $\hat{\theta}(\zeta)$ are real whereas for example 
$e^{i\pi\Delta_1^t\zeta}\theta(\zeta+\int_{p_1}^{p_3}\!\!\omega)=e^{i\pi\Delta_1^t\zeta}\theta(\zeta+\half\Delta_2 + \half \Pi\Delta_1)$ is purely imaginary. 
This explains the minus sign in (\ref{alphasol}) and proves that $\alpha$ is real. The matrix $T$ is constructed \cite{BB} from an involution of the Riemann 
surface that shuffles the basis of cycles $(a_i,b_i)$. To prove for example that $\theta(\zeta)$ is real we use
\beqa
 \bar{\theta}(\zeta)  &=& \sum_{n\in\mathbb{Z}^g} e^{-2\pi i (n^t \bar{\zeta} +\half n^t \bar{\Pi} n)}  \\
                                &=& \sum_{n\in\mathbb{Z}^g} e^{-2\pi i (-n^t T \zeta - \half n^t T \Pi T n)} \\
                                &=& \theta(\zeta)  \ ,
\eeqa
where in the last equation we redefine the summation variable $n\rightarrow Tn$ and used $T^t=T$, $T^2=1$.

\subsection{Solution to equations for $\psi$, $\tilde{\psi}$}

In the previous section we showed in detail how to use the properties of the theta function to solve the cosh-Gordon equation. Now we are going to do the same
for the equations determining $\psi$ but in a more sketchy way. Notice that $\psi$ and $\tilde{\psi}$ are two linearly independent solutions of the same equations:
\beqa
\partial \psi_1 &=&  -\half \partial\alpha \psi_1 + \frac{1}{\lambda} e^{\alpha} \psi_2 \ , \label{p1d}\\
\partial \psi_2 &=& e^{-\alpha} \psi_1 + \half \partial\alpha \psi_2 \ ,  \label{p2d}\\
\bar{\partial} \psi_1 &=&  \half \bar{\partial}\alpha \psi_1 -  e^{-\alpha} \psi_2\ , \label{p1db}\\
\bar{\partial} \psi_2 &=& \frac{1}{\lambda} e^{\alpha} \psi_1 - \half \bar{\partial}\alpha \psi_2 \ ,\label{p2db}  
\eeqa
which are the expanded version of eq.(\ref{p12eq}). 
As a first step we can define a function $F=e^\alpha \frac{\psi_1}{\psi_2}$ which satisfies
\beq
 \partial \ln F = \frac{1}{\lambda} e^{2\alpha} \frac{1}{F} - e^{-2\alpha} F .
 \eeq
 By using the identities for the first derivatives of the theta functions and the value of $e^{2\alpha}$ already given one can readily see that 
 \beq
  F = -\frac{2 D_{p_3}\theta(a) \theta(a+\int_{p_4}^{p_1})}{\theta(a+\int_{p_4}^{p_3})\theta(a+\int_{p_3}^{p_1})} 
  e^{-2\pi i \Delta_1^t\zeta-\frac{i\pi}{2}\Delta_1^t\Pi\Delta_1} 
  \frac{\theta(\zeta)\theta(\zeta+\int_{p_3}^{p_4})}{\theta(\zeta+\int_{p_1}^{p_3})\theta(\zeta+\int_{p_1}^{p_4})}\ ,
\eeq
where we introduced another special point in the Riemann surface that we call $p_4$. The points $p_{1,3}$ are going to be taken as branching points, in particular
for definiteness we take $p_1=0$ and $p_3=\infty$. On the other hand $p_4$ is not a branching point and we take it to be on the upper sheet with $p_4=\lambda$,
the spectral parameter. Going back to the equations for $\psi_{1,2}$ and using the same techniques we find that 
\beqa
\psi_1 &=& C e^{-\half \alpha} \frac{\theta(\zeta+\int_{p_1}^{p_3}+\int_{p_1}^{p_4})}{\theta(\zeta+\int_{p_1}^{p_3})} 
  e^{2 z D_{p_3}\ln\theta(\int_{p_4}^{p_1})+ 2\bar{z} D_{p_1}\ln\theta(a+\int_{p_4}^{p_1})+2\pi i \bar{z} \Delta_1^t\omega(0) }\ ,\nonumber \\
\psi_2 &=&  e^{\half \alpha} \frac{\theta(\zeta+\int_{p_1}^{p_4})}{\theta(\zeta)} 
  e^{2 z D_{p_3}\ln\theta(\int_{p_4}^{p_1})+ 2\bar{z} D_{p_1}\ln\theta(a+\int_{p_4}^{p_1})+2\pi i \bar{z} \Delta_1^t\omega(0)}\ , 
\eeqa
the constant $C$ is determined to be
\beq
C = \frac{2 D_{p_3} \theta(a) \theta(a-\int_{p_1}^{p_4})}{\theta(\int_{p_1}^{p_4})\theta(0)} e^{\frac{i\pi}{2}\Delta_1^t\Pi\Delta_1}.
\eeq
Again, we emphasize that the technique is to match the equation with the properties of the theta functions and choose the parameters appropriately.  
Another, linearly independent solution can be obtained by choosing a different point $p_4$ that we call $\bar{p}_4$. However it has to be associated 
to the same value of $\lambda$ and therefore it can only be the same point but on the other (lower) sheet of the Riemann surface. Namely both $p_4$ 
and $\bar{p}_4$ project on $\lambda$. It should be noticed that, when $p_1=0$, namely one of the branching points, this implies 
\beq
\int_{p_1}^{p_4} \omega_j = - \int_{p_1}^{\bar{p}_4} \omega_j\ ,
\eeq
because the first integral is done on the upper sheet and the second one on the lower sheet where the function $\mu$ changes sign 
(and therefore $\omega_j$ changes sign). We  obtain
\beq
\mathbb{A} = \frac{1}{(\psi_1 \tilde{\psi}_2-\psi_2\tilde{\psi}_1)^{\half}} \left(\begin{array}{cc} \psi_1 & \psi_2 \\ \tilde{\psi}_1 & \tilde{\psi}_2 \end{array}\right).
\eeq
The (constant) normalization factor can be computed using the trisecant identity to give
\beqa
(\psi_1 \tilde{\psi}_2-\psi_2\tilde{\psi}_1)  &=& -2 D_{p_3}\theta(a) e^{\frac{i\pi}{2}\Delta_1^t\Pi\Delta_1} 
\frac{\theta(a+2\int_{p_1}^{p_4})}{\theta^2(\int_{p_1}^{p_4})} e^{2\pi i \Delta_1^t \int_{p_1}^{p_4}} \\
    &=& 2 D_{p_3} \hat{\theta}(0) \frac{\hat{\theta}(2\int_{p_1}^{p_4})}{\theta^2(\int_{p_1}^{p_4})}.
\eeqa
To finish this section we rewrite the solution using the function $\hat{\theta}$ to obtain
\beqa
\psi_1 &=& 2 \frac{D_{p_3}\hat{\theta}(0)}{\theta(0)}\frac{\hat{\theta}(\int_{p_1}^{p_4})}{\theta(\int_{p_1}^{p_4})}
                    \frac{\hat{\theta}(\zeta+\int_{p_1}^{p_4})}{\hat{\theta}(\zeta)} e^{-\half\alpha} e^{\mu z + \nu \bar{z}} \\ &&\nonumber\\
\psi_2 &=& \frac{\theta(\zeta+\int_{p_1}^{p_4})}{\theta(\zeta)} e^{\half \alpha} e^{\mu z + \nu \bar{z}}\ ,
\eeqa
with 
\beq
\mu = -2 D_{p_3} \ln \theta(\int_{p_1}^{p_4}), \ \ \ \ \nu =  -2 D_{p_1} \ln \hat{\theta}(\int_{p_1}^{p_4}) . \label{munudef}
\eeq
It is straight-forward to check directly that these functions satisfy equations (\ref{p1d}),(\ref{p2d}),(\ref{p1db}),(\ref{p2db}).
The only identities that are needed are
\beq
- 4 D_{p_1}\theta(a) D_{p_3}\theta(a) e^{i\frac{\pi}{2}\Delta_1^t\Pi\Delta_1} = 4 D_{p_1}\hat{\theta}(0) D_{p_3}\hat{\theta}(0) = \theta(0)^2\ ,
\eeq 
and
\beq
\lambda = -4 e^{-i\pi\Delta_1^t\Pi\Delta_1+2\pi i \Delta_1^t \int_{p_1}^{p_4}} 
       \left[\frac{D_{p_3}\theta(a) \theta(\int_{p_3}^{p_4})}{\theta(a+\int_{p_4}^{p_3}) \theta(0)}\right]^2 
        = -4 \left[\frac{D_{p_3}\hat{\theta}(0) \hat{\theta}(\int_{p_1}^{p_4})}{\theta(\int_{p_1}^{p_4}) \theta(0)}\right]^2  \ ,          
\eeq
which are explained in the appendix.
The last identity allows us to define (that is to appropriately choose the sign of the square root)
\beq
\sqrt{-\lambda} \equiv 2 \frac{D_{p_3}\hat{\theta}(0) \hat{\theta}(\int_{p_1}^{p_4})}{\theta(\int_{p_1}^{p_4}) \theta(0)}.
\eeq
Then the final form for $\psi_{1,2}$ is simply:
\beqa
\psi_1 &=& \sqrt{-\lambda}\ \frac{\hat{\theta}(\zeta+\int_{p_1}^{p_4})}{\hat{\theta}(\zeta)} e^{-\half\alpha} e^{\mu z + \nu \bar{z}} \\ &&\nonumber\\
\psi_2 &=& \frac{\theta(\zeta+\int_{p_1}^{p_4})}{\theta(\zeta)} e^{\half \alpha} e^{\mu z + \nu \bar{z}}\ ,
\eeqa
where the sign of the square root is chosen according to the previous equation and $\mu$, $\nu$ were defined in eq.(\ref{munudef}). 
We can also compute (remembering that $\int_{p_1}^{p_4}=-\int_{p_1}^{\bar{p}_4}$ because $p_1=0$.)
\beqa
\tilde{\psi}_1 &=& -\sqrt{-\lambda}\ \frac{\hat{\theta}(\zeta-\int_{p_1}^{p_4})}{\hat{\theta}(\zeta)} e^{-\half\alpha} e^{-\mu z - \nu \bar{z}} \ ,\\ &&\nonumber\\
\tilde{\psi}_2 &=& \frac{\theta(\zeta-\int_{p_1}^{p_4})}{\theta(\zeta)} e^{\half \alpha} e^{-\mu z - \nu \bar{z}}.
\eeqa
At this point we can replace $\psi_{1,2}$ and $\tilde{\psi}_{1,2}$ in $\mathbb{A}$ and then in $\mathbb{X}$. This allows us to compute the 
solution directly in Poincare coordinates as:
\beqa
Z &=& \left|\frac{\hat{\theta}(2\int_{p_1}^{p_4})}{\hat{\theta}(\int_{p_1}^{p_4})\theta(\int_{p_1}^{p_4})}\right| 
\frac{|\theta(0)\theta(\zeta)\hat{\theta}(\zeta)|\left|e^{\mu z + \nu \bar{z}}\right|^2}{|\hat{\theta}(\zeta-\int_{p_1}^{p_4})|^2+|\theta(\zeta-\int_{p_1}^{p_4})|^2} \ ,
\label{Zsol}\\
 && \nonumber \\
X+iY &=& e^{2\bar{\mu}\bar{z}+2\bar{\nu}z}\  
  \frac{\theta(\zeta-\int_{p_1}^{p_4})\overline{\theta(\zeta+\int_{p_1}^{p_4})}-\hat{\theta}(\zeta-\int_{p_1}^{p_4})\overline{\hat{\theta}(\zeta+\int_{p_1}^{p_4})}}{|\hat{\theta}(\zeta-\int_{p_1}^{p_4})|^2+|\theta(\zeta-\int_{p_1}^{p_4})|^2}\ , \label{XYsol}
\eeqa

\subsection{Shape of the Wilson loop}

 The shape of the Wilson loop is determined by the intersection of the surface with the boundary.
The boundary is located at $Z=0$ which,  from eq.(\ref{Zsol}) and for finite $z,\bar{z}$ only happens if 
\beq
 Z=0 \ \ \ \Leftrightarrow \ \ \  \theta(\zeta)=0 \ \mbox{or} \ \hat{\theta}(\zeta)=0 .
\eeq
Later on we are going to deal just with the second case so we determine the shape of the Wilson loop by
\beq
\hat{\theta}(\zeta)=0 .
\eeq
This equation defines a curve in the world-sheet which in turn is mapped to a curve in the 
$(X,Y)$ plane using the solution to the equations of motion (\ref{XYsol}).

\subsection{Computation of the Area}

 The expectation value of the Wilson loop is determined by the area of the minimal surface we described.
In conformal gauge the area is computed as:
\beq
 A = 2 \int \partial X_\mu \bar{\partial}X^\mu d\sigma d\tau = 2 \int \Lambda d\sigma d\tau 
= 4 \int e^{2\alpha} d\sigma d\tau \ ,
\label{Areadef}
\eeq
where we used eqns.(\ref{Lcomp}) to write the area in terms of the Lagrange multiplier $\Lambda$ and then used that
\beq
\Lambda=2e^{2\alpha}\ ,
\eeq
as follows from eqns.(\ref{LJ}) and (\ref{JJbar}) using that $|\lambda|=1$.
 We could in principle replace $\alpha$ by its expression in eq.(\ref{alphasol}) and evaluate the integral 
numerically but such procedure fails because the area is divergent. The correct procedure is to identify the divergent 
piece analytically and then extract the finite piece in terms of a finite integral that can be easily evaluated 
numerically. In order to do so we use eqs. (\ref{alphasol}) and (\ref{Thetapp}) (using $a=\int_{p_1}^{p_3}$) to get:
\beqa
e^{2\alpha} &=& 4 \left\{ D_{p_1p_3}\ln \theta(0)- D_{p_1p_3}\ln \hat{\theta}(\zeta)\right\} \\
           &=& 4 D_{p_1p_3}\ln \theta(0) - \partial\bar{\partial}\ln \hat{\theta}(\zeta)  . \label{alphaD13}
\eeqa
The first term in the expression for $e^{2\alpha}$ is a constant and the second one is a total derivative. 
The first integral is clearly finite but the second one contains the divergent piece that we need to regulate.
In order to do that we observe that for these solutions
\beq
 Z = |\hat{\theta}(\zeta)| h(z,\bar{z})\ ,
\eeq
where $Z$ is one of the Poincare coordinates and
\beq
h(z,\bar{z})=\left|\frac{\hat{\theta}(2\int_{p_1}^{p_4})}{\hat{\theta}(\int_{p_1}^{p_4})\theta(\int_{p_1}^{p_4})}\right| 
\frac{|\theta(0)\theta(\zeta)|\left|e^{\mu z + \nu \bar{z}}\right|^2}{|\hat{\theta}(\zeta-\int_{p_1}^{p_4})|^2+|\theta(\zeta-\int_{p_1}^{p_4})|^2} .
\eeq
 Furthermore using Stokes or Gauss theorem we find that for any well-behaved 
function $F$:
\beq
 \int d\sigma d\tau \partial \bar{\partial} F = \frac{1}{4} \int d\sigma d\tau  \nabla^2 F
 = \frac{1}{4} \oint \hat{n}\cdot\nabla F d\ell \ ,
\eeq
 where the contour integral is over the boundary of the Wilson loop (in the world-sheet), $\hat{n}$ is an outgoing
normal vector and $d\ell$ is the differential of arc length. 
The area is then:
\beq
A = 16 D_{p_1p_3}\ln \theta(0) \int d\sigma d\tau +  \oint \hat{n}\cdot \nabla\ln h \ d\ell
    -  \oint \hat{n}\cdot \nabla \ln Z\ d\ell .
\eeq
The last integral is divergent and we concentrate now on extracting the leading divergence. The correct AdS/CFT prescription
is to cut the surface at $Z=\epsilon$ and write the area as
\beq
 A = \frac{L}{\epsilon} + A_{f}\ ,
\eeq
 where $L$ should be the length of the Wilson loop and $A_f$ is the finite part which is identified with the expectation value of the Wilson loop
 through:
 \beq
  \langle W \rangle = e^{-\frac{\sqrt{\lambda}}{2\pi} A_f}\ ,
 \eeq 
 where here $\lambda$ is the 't Hooft coupling of the gauge theory (not to be confused with the spectral parameter). This prescription is 
 equivalent to subtracting the area $A=\frac{L}{\epsilon}$ of a string ending on the contour of length $L$ and stretching along $Z$ from the boundary 
 to the horizon.  To see that the coefficient of the divergence is indeed the length, let us compute
\beq
 A_{\mbox{div.}} =  -  \oint_{Z=\epsilon} \frac{1}{Z} \hat{n}\cdot \nabla Z  d\ell
                =  \frac{1}{\epsilon} \oint_{Z=\epsilon}  |\nabla Z| d\ell\ ,
\eeq
where we used that the normal is precisely in the opposite direction of $\nabla Z$ because the contour is a curve of constant 
$Z=\epsilon$ and $Z$ increases toward the inside. On the other hand the length in the boundary is given by
\beq
 L = \oint \sqrt{|\hat{t}.\nabla X|^2 + |\hat{t}.\nabla Y|^2} d\ell \ ,
\eeq
where $\hat{t}$ is a unit vector tangent to the contour. We can move forward if we write the equation of motion for $X$ as derived form the 
action (\ref{action}):
\beq
2 \nabla X \cdot \nabla Z =  Z \nabla^2 X\ ,
\eeq
which, when $Z\rightarrow 0$, becomes $\nabla X\cdot \nabla Z=0$ namely $\nabla X$ is perpendicular to the normal and 
therefore parallel to the tangent $\hat{t}$. The same is true for $Y$ so we find
\beq
 L = \oint \sqrt{|\nabla X|^2 + |\nabla Y|^2} d\ell + \cO(\epsilon^2) .
\eeq
Finally the equation of motion for $Z$ is
\beq
 (\nabla Z)^2 - Z \nabla^2 Z = (\nabla X)^2 + (\nabla Y)^2\ ,
\eeq
which for $Z\rightarrow 0$ implies that $\sqrt{|\nabla X|^2 + |\nabla Y|^2}=|\nabla Z|$.
Therefore the length of the Wilson loop is given by
\beq 
 L = \oint |\nabla Z| d\ell - \frac{\epsilon}{2} \oint \frac{\nabla^2 Z}{|\nabla Z|}\ d\ell \ ,
\eeq
and the divergent piece of the area is indeed $A_{\mbox{div.}}=\frac{L}{\epsilon}$. There is a finite part remaining:
\beqa
A &=& \frac{L}{\epsilon} + A_{\mbox{f}} \ ,\\
A_{\mbox{f}}  &=& 
 16 D_{p_1p_3}\ln \theta(0) \int d\sigma d\tau +  \oint \hat{n}\cdot \nabla\ln h \ d\ell + \half \oint \frac{\nabla^2 Z}{|\nabla Z|}\ d\ell . \nonumber
 \eeqa
 The integrals are performed on the world-sheet parameterized by $\sigma$, $\tau$. The first integral is proportional to the area of the world-sheet. The last two
 integrals are done over the world-sheet boundary. The final expression can be simplified by rewriting $Z = |\hat{\theta}(\zeta)| h(z,\bar{z})$ and using that
 $\hat{\theta}(\zeta)$ vanishes on the boundary where the contour integral is performed. It is then easy to check that $h(z,\bar{z})$ cancels and the final 
 formula for the area is:
\beqa
A &=& \frac{L}{\epsilon} + A_{\mbox{f}} \ ,\\
A_{\mbox{f}}  &=& 
 16 D_{p_1p_3}\ln \theta(0) \int d\sigma d\tau - \half \oint \frac{\nabla^2 \hat{\theta}(\zeta)}{|\nabla \hat{\theta}(\zeta)|}\ d\ell \\
 &=& 
 16 D_{p_1p_3}\ln \theta(0) \int d\sigma d\tau - 2 \oint \frac{D_{p_1p_3} \hat{\theta}(\zeta)}{|D_{p_1} \hat{\theta}(\zeta)|}\ d\ell  \label{Afres} \\
 &=& 
 8 D_{p_1p_3}\ln \theta(0) \oint (\sigma d\tau-\tau d\sigma) 
 - 2 \oint \frac{D_{p_1p_3} \hat{\theta}(\zeta)}{|D_{p_1} \hat{\theta}(\zeta)|}\ d\ell. 
 \eeqa
where in the last step we used the well-known formula for the area of the region encircled by a given curve. The renormalized area is then expressed as a 
finite one dimensional contour integral over the boundary of the world-sheet. Perhaps it would also be useful to clarify that 
$|\nabla \hat{\theta}(\zeta)|$ denotes the norm of a real 2-vector whereas  $|D_{p_1} \hat{\theta}(\zeta)|$  is the modulus of a complex number. 
The final expression for the area is quite interesting because it does not depend on the spectral parameter $\lambda$. 
Therefore the shape of both, the Wilson loop and the dual surface depend on the parameter $\lambda$ but the area $A_f$ does not. 
In this way we explicitly find a one parameter family of deformations that preserve the area. It is not obvious at
first that the area $A_f$ should be independent of the spectral parameter because, although the definition (\ref{Areadef}) does not contain $\lambda$, 
the regularized  area does since, as we said, $L$ depends on $\lambda$. It so happens that the finite part $A_f$ does not.  
The situation is similar to scale transformations that  modify $L$ but not $A_f$.
 
\begin{figure}
\centering
\includegraphics[width=10cm]{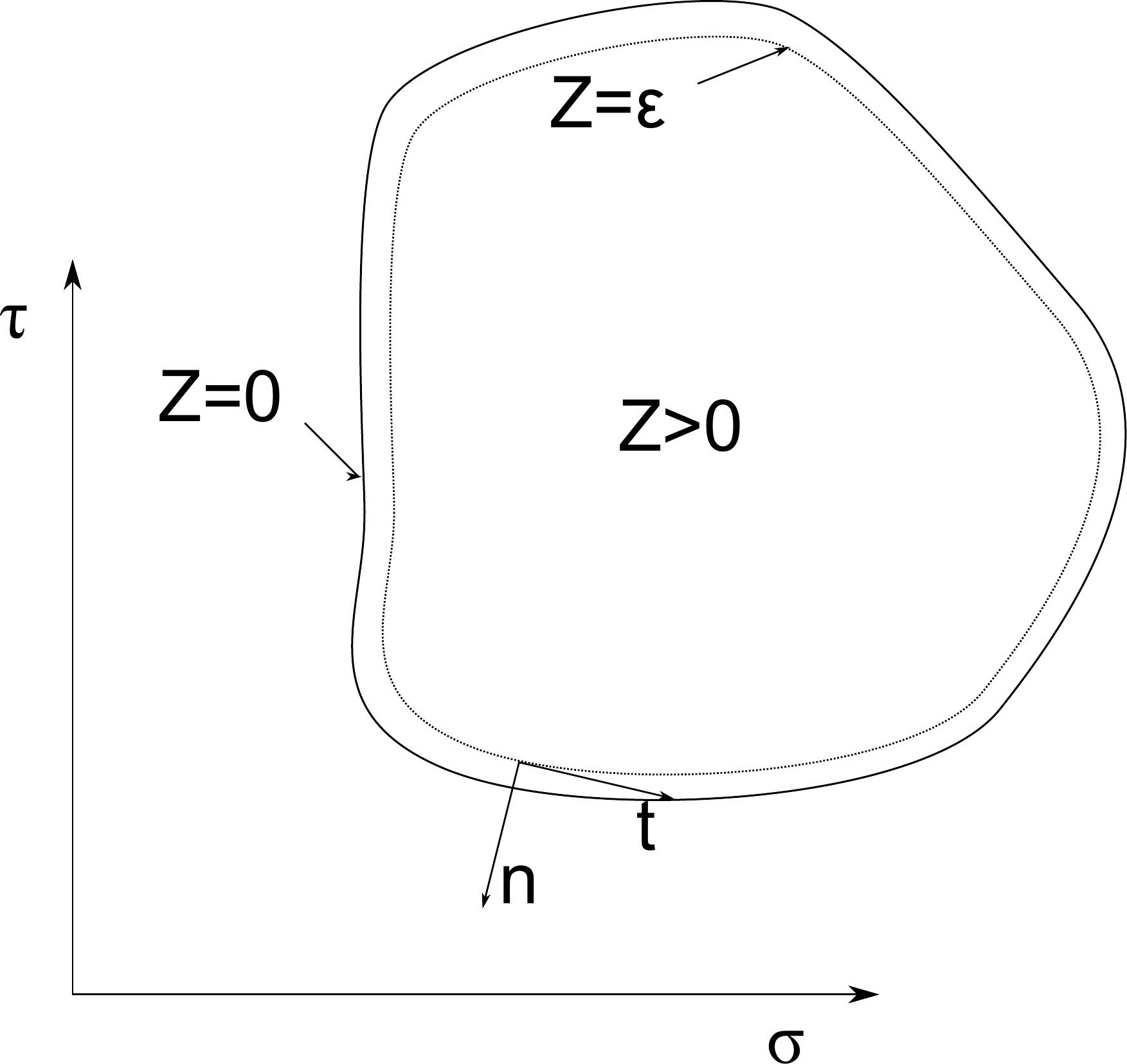}
\caption{The boundary is determined by the contour $Z=0$. However the area is computed by integrating up to a
contour $Z=\epsilon\rightarrow 0$ and then the leading divergence $\frac{L}{\epsilon}$ is subtracted. Here $L$ is 
the length of the contour in the boundary (not in this ($\sigma$,$\tau$) plane).}
\label{contour}
\end{figure}

\section{Example with $g=3$}

We are going to consider an example to illustrate the shape of the Wilson loops that are obtained in this way. The main purpose is to show that we find closed 
Wilson loops whose dual surface is known analytically. 

\begin{figure}
\centering
\includegraphics[width=12cm]{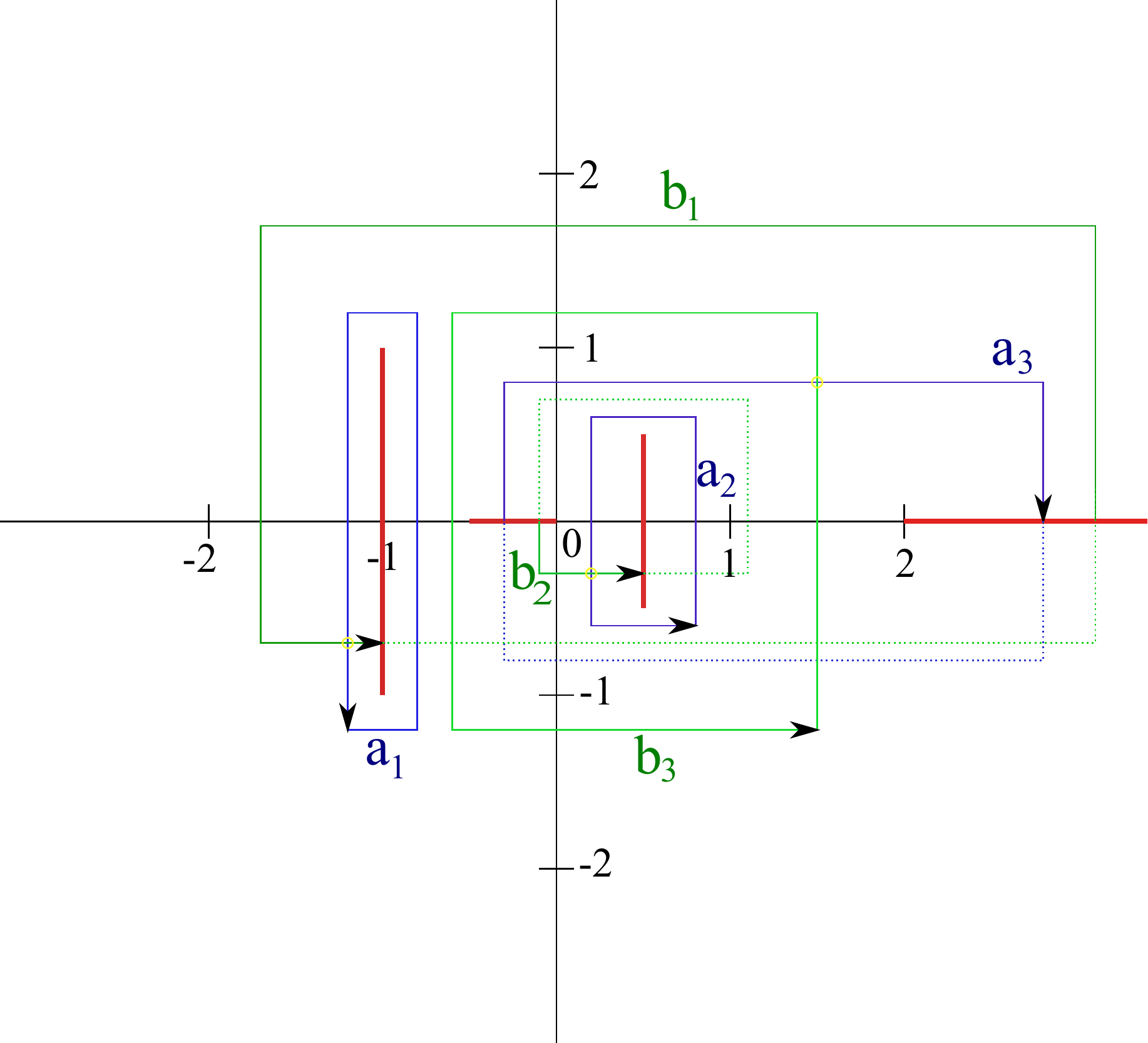}
\caption{Genus three Riemann surface. A basis of fundamental cycles is depicted with solid paths being in the upper sheet and dotted lines in the lower 
sheet. The circles show their intersection points. }
\label{cuts}
\end{figure}

Consider the function 
\beq
\mu = i \sqrt{-i(\lambda+1-i)} \sqrt{-i(\lambda+1+i)} \sqrt{-i(\lambda-\frac{1+i}{2})} \sqrt{-i(\lambda-\frac{1-i}{2})} 
       \sqrt{2-\lambda} \sqrt{\lambda} \sqrt{\lambda+\half}\ ,
\eeq
where the square root is taken to have a cut in the negative real axis. So defined, the function $\mu$ has cuts in the complex plane as illustrated in fig.\ref{cuts}
but is smooth in a double cover of the plane which defines a hyperelliptic Riemann surface of genus $g=3$. The cycles $a_i$, $b_i$ are taken as in the 
figure. The Riemann surface has an involution $\lambda\rightarrow -\frac{1}{\bar{\lambda}}$ meaning that knowing the cuts for $|\lambda|>1$ we can
reconstruct the cuts inside the unit circle. This involution is important to construct the matrix $T$ that fixes the correct reality conditions (see \cite{BB}).  
A basis for the holomorphic abelian differentials is given by 
\beq
 \nu_k = \frac{\lambda^{k-1}}{\mu} d\lambda, \ \ \ \ k=1\ldots 3 .
\eeq
 If we compute
\beq
C_{ij} = \oint_{a_i} \nu_j , \ \ \ \ \tilde{C}_{ij} = \oint_{b_i} \nu_j\ ,
\eeq
then a normalized basis of holomorphic abelian differentials is
\beq
\omega_i = \nu_j \left(C^{-1}\right)_{ji}\ ,
\label{wdef}
\eeq
and the period matrix is
\beq
\Pi = \tilde{C} C^{-1} =\left( \begin{array}{ccc}0.5+0.64972 i&0.14972 i&-0.5\\ 0.14972 i&-0.5+0.64972 i&0.5\\ -0.5&0.5&0.639631\end{array}\right) .
\eeq
The Jacobi map (with base at 0) is defined as
\beq
\phi(\lambda)_j = \int_0^{\lambda} \omega_j = \int_0^\lambda \frac{\lambda^{j-1}}{\mu} \left(C^{-1}\right)_{ji} d\lambda .
\eeq
The function $\theta(\phi(\lambda))$ has three zeros: $\lambda=\infty, \frac{1-i}{2}, -1+i$. To prove this we can take, on the upper sheet, 
a path from $\lambda=0$ to each of the zeros of $\mu$. Coming back along the lower sheet defines a closed path $\mathcal{C}_\lambda$
equivalent to:
\beq
\begin{array}{lclcl}
\lambda= -\half           &\ \ \rightarrow\ \ & \mathcal{C}_\lambda \equiv b_3-a_2              &\ \ \rightarrow\ \ & 
                                                                      \theta\left[  \begin{array}{ccc}  0&0&1\\0&-1&0 \end{array} \right]\!(\zeta) \\
\lambda= \half - \half i  &    \rightarrow    & \mathcal{C}_\lambda \equiv a_2+b_2              &    \rightarrow    & 
                                                                      \theta\left[ \begin{array}{ccc} 0&1&0\\0&1&0 \end{array}\right]\!(\zeta)  \\
\lambda= \half + \half i  &    \rightarrow    & \mathcal{C}_\lambda \equiv b_2                  &    \rightarrow    & 
                                                                      \theta\left[ \begin{array}{ccc} 0&1&0\\0&0&0 \end{array}\right]\!(\zeta)  \\
\lambda= 2                &    \rightarrow    & \mathcal{C}_\lambda \equiv a_3-a_2              &    \rightarrow    & 
                                                                      \theta\left[ \begin{array}{ccc} 0&0&0\\0&-1&1 \end{array}\right]\!(\zeta) \label{Ch} \\
\lambda= -1+i             &    \rightarrow    & \mathcal{C}_\lambda \equiv -a_2+a_3+b_1+b_3     &    \rightarrow    & 
                                                                      \theta\left[ \begin{array}{ccc} 1&0&1\\0&-1&1 \end{array}\right]\!(\zeta) \\
\lambda= -1-i             &    \rightarrow    & \mathcal{C}_\lambda \equiv -a_1+a_2+a_3+b_1-b_3 &    \rightarrow    & 
                                                                      \theta\left[ \begin{array}{ccc} 1&0&-1\\-1&1&1 \end{array}\right]\!(\zeta)\\
\lambda= \infty           &    \rightarrow    & \mathcal{C}_\lambda \equiv a_1-a_2+a_3+b_3      &    \rightarrow    & 
                                                                      \theta\left[ \begin{array}{ccc} 0&0&1\\1&-1&1 \end{array}\right]\!(\zeta) .
\end{array}
\eeq
The last column gives a theta function with characteristic determined by $\left[ \begin{array}{c}\varepsilon_1 \\ \varepsilon_2 \end{array} \right]$ where 
$\varepsilon_{1,2}$ are given by $\mathcal{C}_\lambda \equiv \sum_{i=1}^3 (\varepsilon_{1i} b_i + \varepsilon_{2i} a_i)$. 
The point is that this theta function vanishes if and only if $\theta(\phi(\lambda))$ vanishes because $\phi(\lambda) = \half(\varepsilon_2 + \Pi \varepsilon_1)$
(since the integrals on the upper and lower sheet are equal to half the total integral).  
 If the characteristic is odd (namely $\varepsilon_1^t \varepsilon_2$ is odd) then by symmetry the theta function is zero at the origin and therefore the
theta function without characteristic is zero at the corresponding point. 

The vector of Riemann constants \cite{FK} is computed from the sum of the characteristic of the zeros which is (mod 2)
\beq
\kappa \equiv \left[ \begin{array}{ccc} 1&1&0\\1&1&0 \end{array} \right]\ ,
\eeq
and has the property that for any $\lambda_{1,2}$ on the Riemann surface we have
\beq
\theta\left[ \begin{array}{ccc} 1&1&0\\1&1&0 \end{array} \right]\left(\phi(\lambda_1)+\phi(\lambda_2)\right)=0 . \label{kappadef}
\eeq
Moreover, this completely defines the set of zeros of the theta functions we are considering \cite{FK}.

\begin{figure}
\centering
\includegraphics[width=10cm]{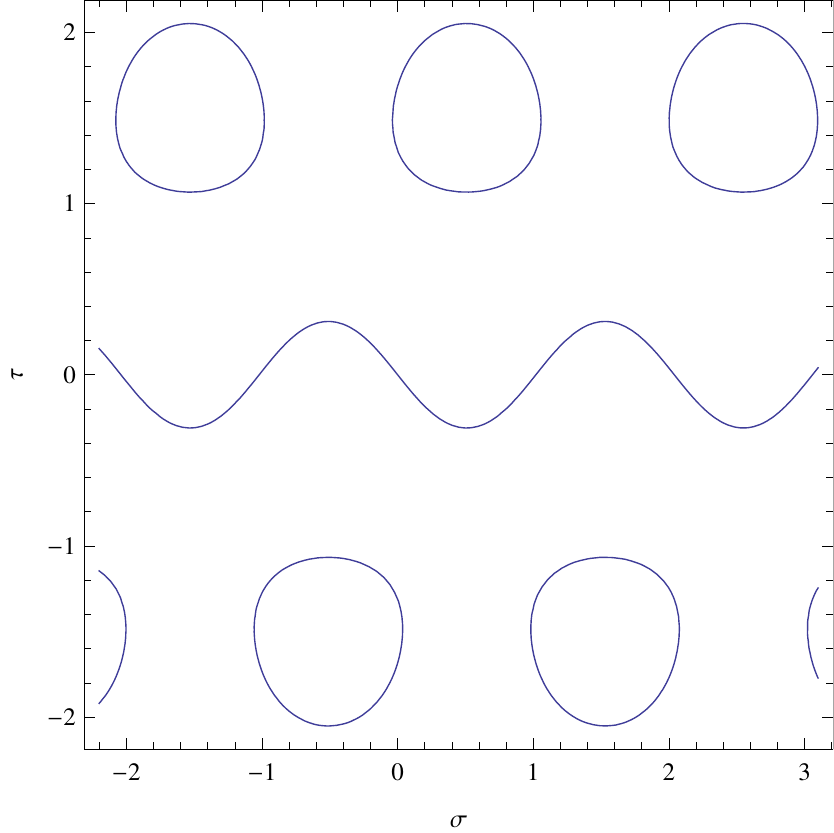}
\caption{Zeros of the function $\hat{\theta}$. We can see the (quasi)-periodicity of the theta function as well as closed contours that map into closed Wilson loops in the boundary. }
\label{zeros_theta_hat}
\end{figure}

 To write the solution we  choose the points $p_1=0$ and $p_3=\infty$ which works well since, as seen above a path between 0 and $\infty$ defines
 and odd characteristic that can be taken to be $\Delta_1=\left[\begin{array}{c}0\\0\\1\end{array}\right]$, 
 $\Delta_2=\left[\begin{array}{c}1\\1\\1\end{array}\right]$. We therefore define
 \beq
 \hat{\theta}(\zeta)= \theta\left[\begin{array}{ccc} 0&0&1 \\ 1&1&1 \end{array}\right](\zeta) .\label{thetahatdef}
 \eeq
 We can now write
 \beq
  \zeta= 2 ( \omega(\infty) z + \omega(0) \bar{z}) = \left[\begin{array}{c} -0.4903\sigma-0.19069\tau\\  -0.4903\sigma+0.19069\tau\\ 0.59321\tau \end{array}\right] \ ,
 \eeq
 where we defined $z=\sigma+i\tau$.
 The zeros of the function $\hat{\theta}$ in the complex plane $z$ determine the boundary and therefore the shape of the Wilson loops. We plot the contours
 where $\hat{\theta}$ becomes zero in fig.\ref{zeros_theta_hat}. 
 
\begin{figure}
\centering
\includegraphics[width=10cm]{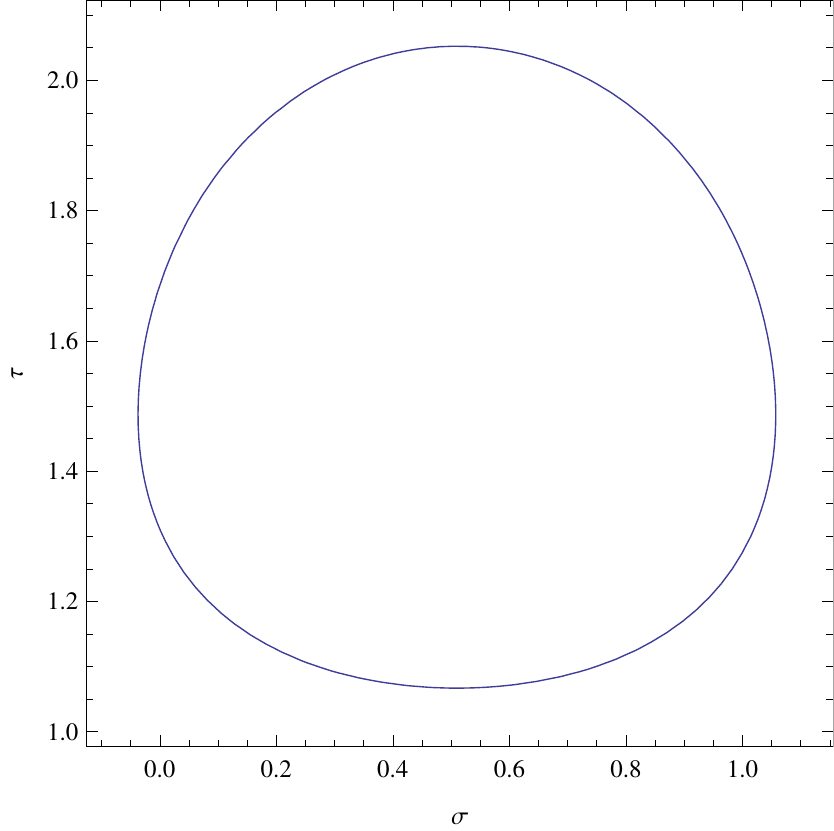}
\caption{Particular contour (taken from fig.\ref{zeros_theta_hat}) in the $z$ plane chosen to compute the minimal area surface.}
\label{zeros_theta_hat_1}
\end{figure}
 
\begin{figure}
\centering
\centering
\subfloat[$\lambda=i$]{\includegraphics[width=6cm]{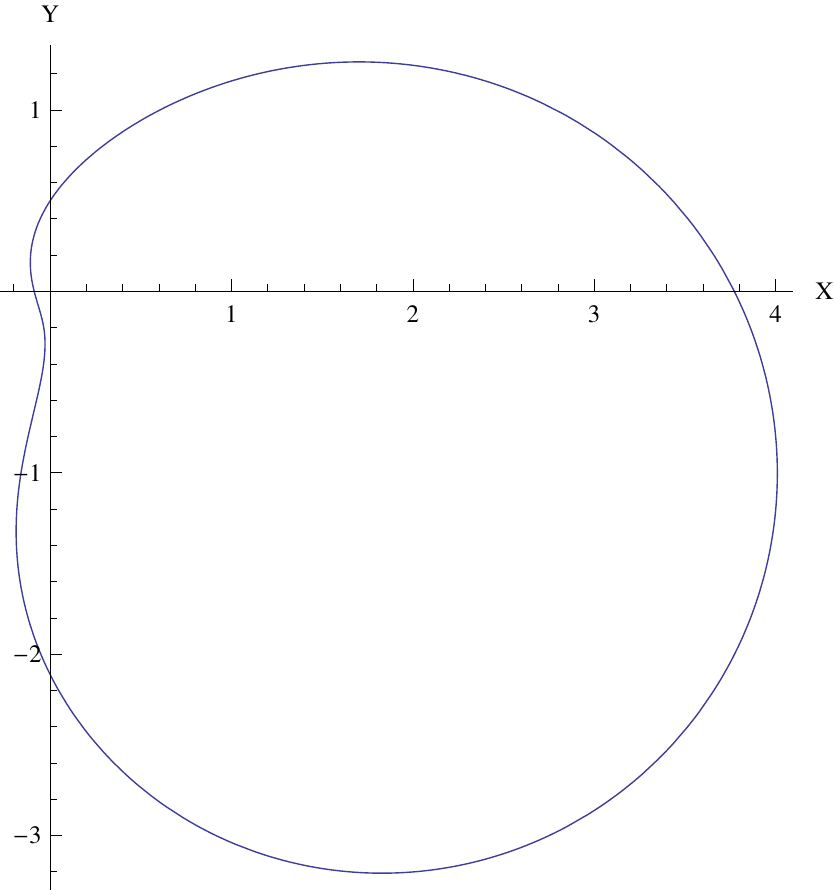}}
\subfloat[$\lambda=-\frac{1+i}{\sqrt{2}}$]{\includegraphics[width=5.5cm]{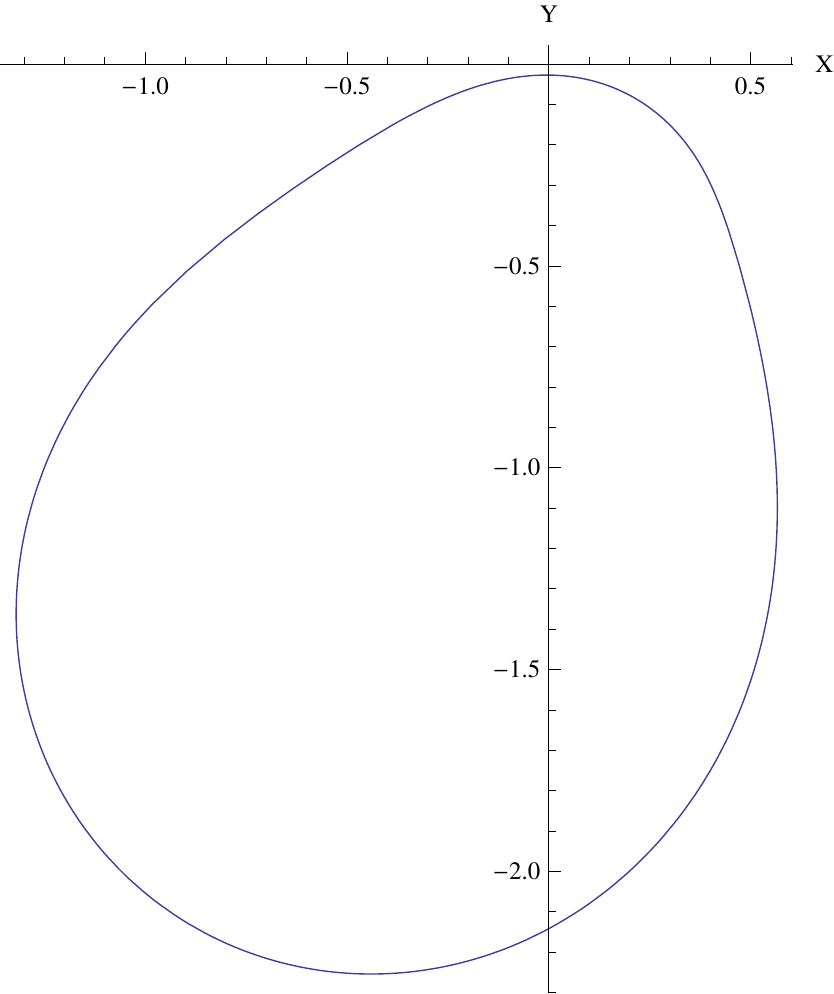}}
\caption{Shape of the Wilson loops that we use as an example. It is obtained by mapping the contour in fig.\ref{zeros_theta_hat_1} into the $X,Y$ plane
for two values of the spectral parameter $\lambda$. }
\label{WL_shape}
\end{figure}

\begin{figure}
\centering
\subfloat[$\lambda=i$]{\includegraphics[width=7.5cm]{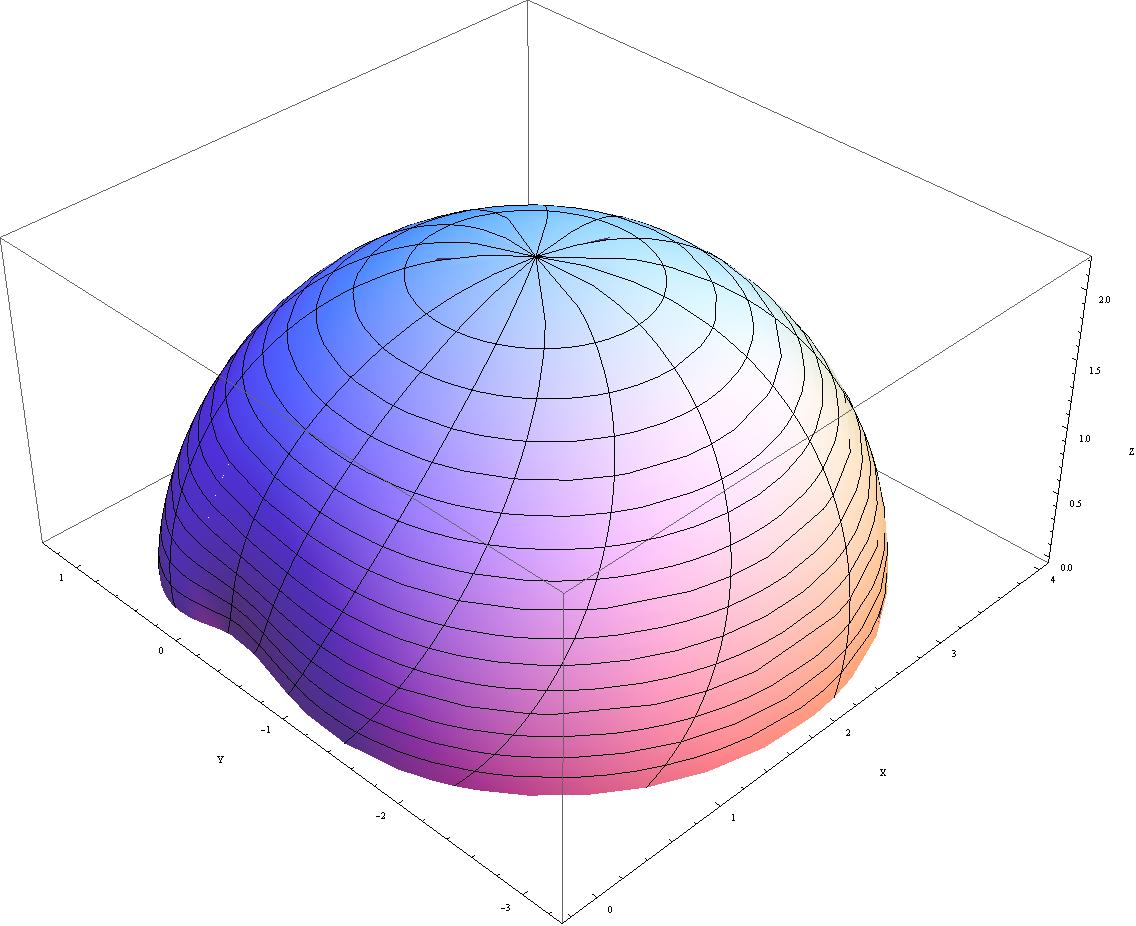}}
\subfloat[$\lambda=-\frac{1+i}{\sqrt{2}}$]{\includegraphics[width=7cm]{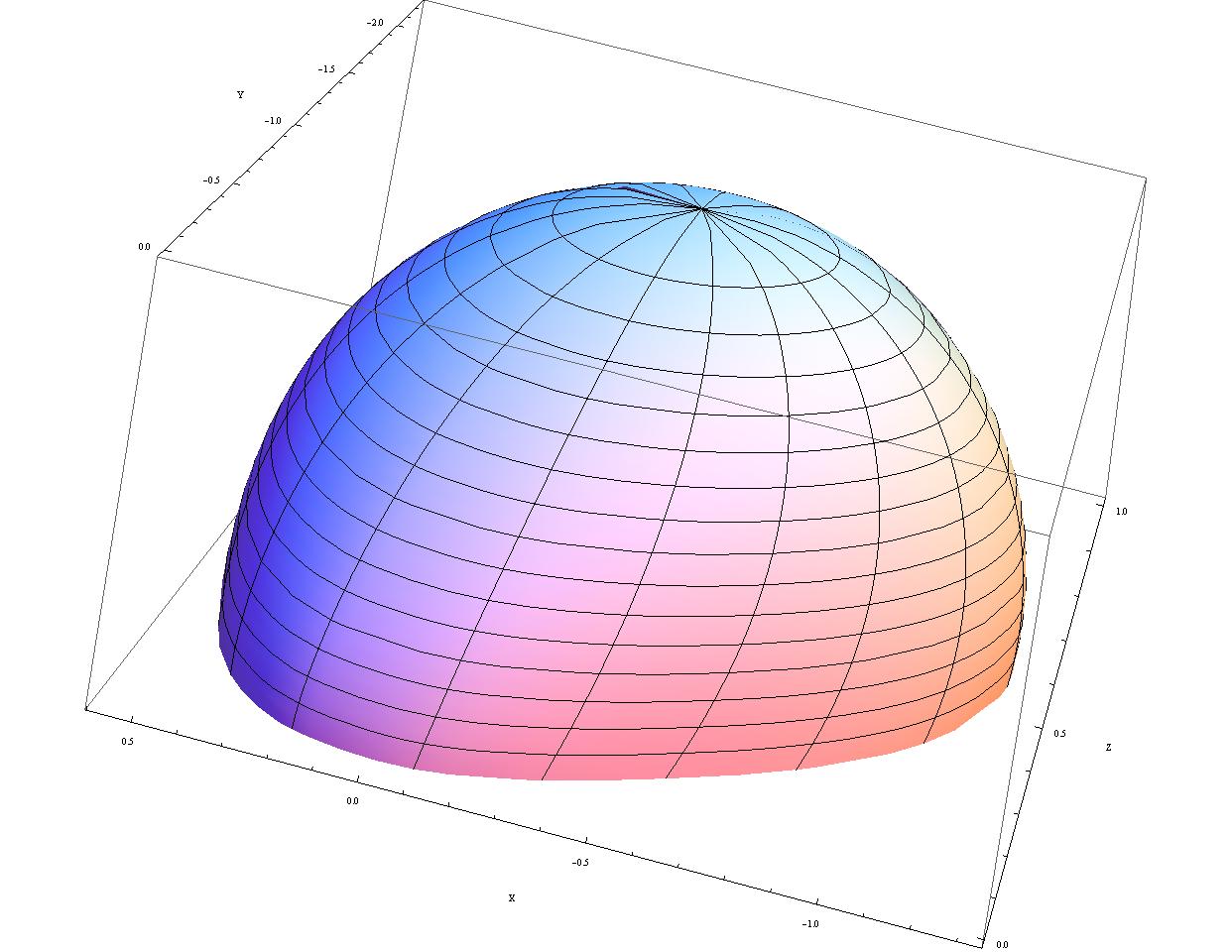}}
\caption{Minimal area surfaces ending on the contours illustrated in fig.\ref{WL_shape}.  We emphasize that the surfaces are known analytically.}
\label{WL_surface}
\end{figure}

 The (quasi)-periodicity of the theta function is evident from the figure but also the existence
 of closed curves which in turn will give rise to closed Wilson loops. Choosing the contour displayed in fig.\ref{zeros_theta_hat_1} determines a Wilson loop up 
 to the spectral parameter $\lambda$ that is still arbitrary. Taking into account that $|\lambda|=1$ we choose as examples
 \beq
 \lambda_1=i, \ \ \ \ \ \lambda_2=-\frac{1+i}{\sqrt{2}} .
 \eeq
 Therefore the point $p_4=i$ (or $p_4=-\frac{1+i}{\sqrt{2}}$) in the formulas for the surface (understood to be in the upper sheet). The shape of the 
 Wilson loop can be immediately obtained by mapping the contour into the  $X,Y$ plane. The result in displayed in fig.\ref{WL_shape}
 where we see a non-trivial closed Wilson loop. The minimal surfaces ending on these contours are displayed in fig.\ref{WL_surface}. 
 Finally we can compute the length of the Wilson loop and the finite part of the area using eq.(\ref{Afres}). The result is
 \beqa
 L_1 &=&  13.901, \ \ \ L_2 = 6.449 \ ,\\  
 A_f  &=& -6.598 \ \ \ \mbox{for both.}
 \eeqa
  The length $L_{1,2}$ can be changed by a scale transformation but the finite part $A_f$ is scale invariant and independent of $\lambda$, the spectral parameter. 
 For comparison, for a circle of radius $R$ the corresponding values
  are:
 \beq
  L^{(circle)} = 2\pi R , \ \ \ A_f^{(circle)} = -2\pi.
 \eeq  
 Although in the end the results are numerical, we emphasize that the shape of the surface is known analytically. Also a relative simple expression 
was found for the expectation value (after subtracting the infinities) in terms of a one-dimensional finite integral. At the end, the integral was 
evaluated numerically to get the value for the area. Notice also that the area is smaller than the one for the circle. For a given length $L$  
the area is not bounded from below since we can take a contour made out of two parallel lines of length $L/2$ and separated by a distance 
$\delta\rightarrow 0$ in which case the area goes to minus infinity as $A_f \sim -\frac{L}{2\delta}$ \cite{MRY}.  
On the other hand, for fixed $L$, the circle is expected to be an upper bound as shown in \cite{SARM}. Our result agrees with that bound.  

\begin{figure}
\centering
\includegraphics[width=10cm]{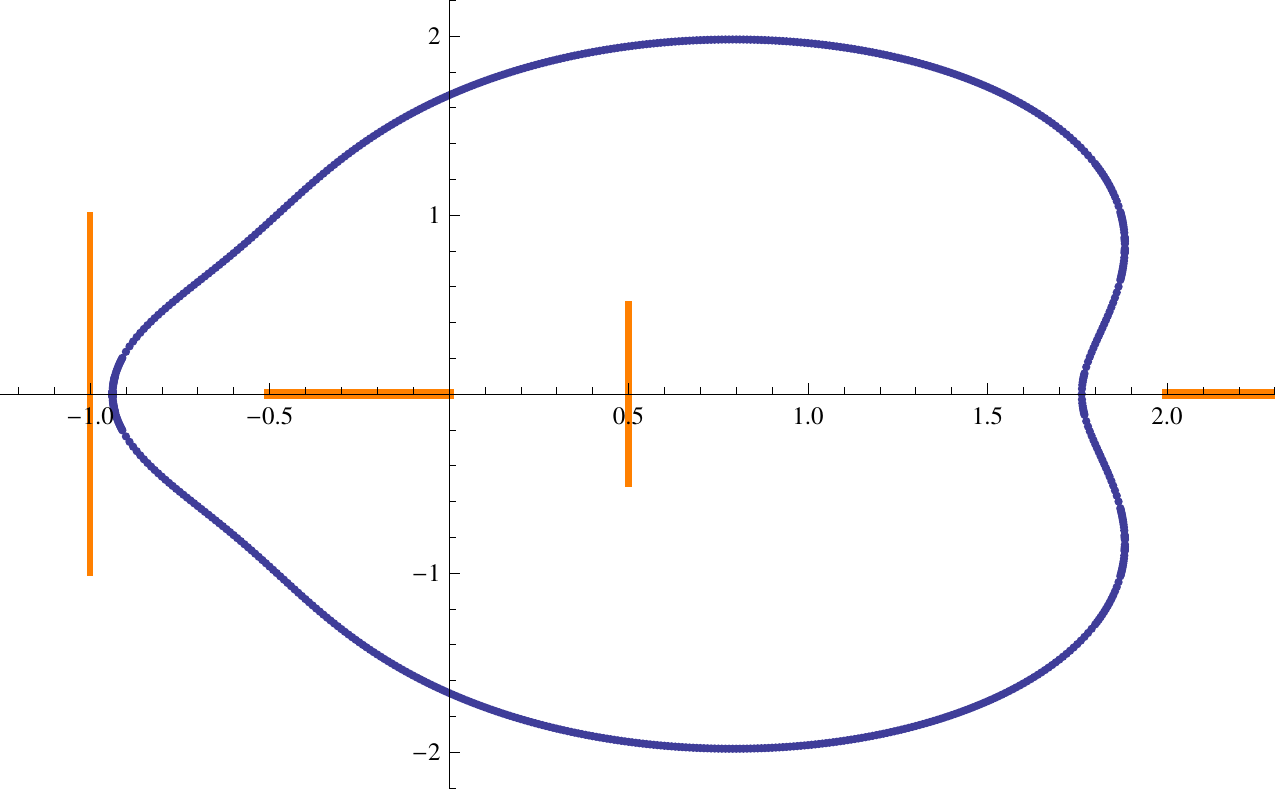}
\caption{For genus three, each Wilson loop computable in terms of theta functions maps into a closed curve inside the corresponding Riemann surface. In this
figure we display the closed curve corresponding to the Wilson loop of fig.\ref{WL_shape}. We notice it encircles two cuts.}
\label{WL_in_RS}
\end{figure}

 Having described a particular example in detail we want to elaborate further on the properties of these Wilson loops. We know that the zeros of $\hat{\theta}(\zeta)$ determine the boundary of the Wilson loop, but, from eqs.(\ref{kappadef}), (\ref{thetahatdef}) we know that all zeros are given by 
\beq
  \zeta = \half\left[\begin{array}{c} 0\\ 0 \\1  \end{array}\right] +\half \Pi\left[\begin{array}{c} 1\\ 1 \\ 1  \end{array}\right]+ \phi(\lambda_1)+\phi(\lambda_2)\ ,
\eeq
for arbitrary points $\lambda_{1,2}$ on the Riemann surface. The vectors  $\left[\begin{array}{c} 0\\ 0 \\1  \end{array}\right] $ and 
$\left[\begin{array}{c} 1\\ 1 \\ 1  \end{array}\right]$ represent the difference (mod 2) between the characteristics of $\hat{\theta}(\zeta)=\theta\left[\begin{array}{ccc} 0&0&1 \\ 1&1&1 \end{array}\right](\zeta)$  and $\kappa \equiv \left[ \begin{array}{ccc} 1&1&0\\1&1&0 \end{array} \right]$.
We cannot take random points $\lambda_{1,2}$ on the Riemann surface because we also have 
\beq
\zeta= 2 \left( \omega_\infty z + \omega_0 \bar{z} \right)\ ,
\eeq
which satisfies
\beq
 \bar{\zeta} =-T\zeta, \ \ \ \ \ T=\left(\begin{array}{ccc}0&-1&0\\-1&0&0\\0&0&-1\end{array}\right) .
 \eeq
  With some work it can be seen that we can take
\beq
 \zeta =\half\left[\begin{array}{c} 2\, n_1\\ 2\, n_2 \\1+2\, n_3  \end{array}\right] +\half \Pi\left[\begin{array}{c} 1+2\, m_1\\ 1+2\, m_2 \\ 1+2\, m_3  \end{array}\right]- \phi(\lambda_1)+\phi\left(-\frac{1}{\bar{\lambda}_1}\right) , \label{iRs}
\eeq
where $n_{1,2,3}$ and $m_{1,2,3}$ are integers which can be absorbed in the definition of the path used to compute the function $\phi$.
 It therefore follows that for genus 3 we can map the Wilson loop into a curve inside the Riemann surface. Namely for each point in the contour displayed in 
 fig.\ref{zeros_theta_hat_1} there is a point $\lambda_1$ in the Riemann surface such that:
 \beq
 2 \left( \omega(\infty) z + \omega(0) \bar{z} \right)=\half\left[\begin{array}{c} 2\, n_1\\ 2\, n_2 \\1+2\, n_3  \end{array}\right] +\half \Pi\left[\begin{array}{c} 1+2\, m_1\\ 1+2\, m_2 \\ 1+2\, m_3  \end{array}\right]- \phi(\lambda_1)+\phi\left(-\frac{1}{\bar{\lambda}_1}\right) . \label{iRs2}
 \eeq
The set of such points describes a curve inside the Riemann surface which is depicted in fig.\ref{WL_in_RS}. 
The statement is the following: for each point $\lambda_1$ in the curve and for a given choice of path used to define the function $\phi$ there is a set of integers 
$n_{1,2,3}$ and $m_{1,2,3}$ such that eq.(\ref{iRs2}) can be solved for $z$. These values of $z$ lie in the closed curves depicted in fig.\ref{zeros_theta_hat}
which are the zeros of the function $\hat{\theta}$. Equivalently we can set all integers $n_{1,2,3}=0$ and $m_{1,2,3}=0$ but then we have to choose the
path used to define $\phi$ appropriately so that eq.(\ref{iRs2}) has a solution. Furthermore, with an appropriate choice we can map the curve in fig.\ref{WL_in_RS}
to the curve in fig.\ref{zeros_theta_hat_1} which was the one used in our examples. 
 For higher genus, there should be a set of curves in the Riemann surface for each closed
 Wilson loop. In this paper we do not explore this issue further but we believe it should be an interesting subject to pursue.
 
As a final comment, it should be noted that these Wilson loops are not BPS since, for Euclidean Wilson loops with constant scalar, the only BPS ones are 
straight lines  \cite{Zarembo:2002an}.

\section{Conclusions} 
\label{conclusions}

 In this paper we discussed minimal area surfaces in \ads{3} space which are dual to Wilson loops in \N{4} SYM.  We essentially follow the results of the paper
 \cite{BB} where such solutions were found but provide a different derivation of the solutions and also a formula for the finite part of the area in accordance
 with the one used in the AdS/CFT correspondence.  Finally we make the observation that closed Wilson loops appear from these solutions in which case the 
 world-sheet has the topology of a disk and the area is expressed as a (finite) contour integral over the world-sheet boundary.
 In this way we construct an infinite parameter family of new examples of Wilson loops whose dual surface is analytically known. 
 It should be noticed that, to our knowledge, only the circular Wilson loop and the lens shaped loop were previously known for closed Euclidean Wilson loop (with constant scalar). Furthermore, the result gives, for each individual Wilson loop a one parameter family
 of deformation (given by the spectral parameter $\lambda$) such that the area remains the same. 
 Finally, for genus 3 we pointed out an interesting map between Wilson loops computable in this way and curves inside the Riemann surface.  
  We hope that these new solutions will give rise to a better understanding of Wilson loops in the context of the AdS/CFT correspondence and also in general. 
  It is evident that an important integrable structure lies behind them that should be explored in detail. It would be interesting if these ideas allow us to reconstruct 
  the shape of the Wilson loop from the field theory using some sort of coherent states formalism as in \cite{spinchain}.   
 
\section{Acknowledgments}

We are grateful to Nadav Drukker, Georgios Michalogiorgakis, Alin Tirziu and Peter Ouyang for comments and suggestions.
This work was supported in part by NSF through grants PHY-0805948, a CAREER Award PHY-0952630, a Graduate Fellowship (S.Z.) and an AGEP 
grant \#0450373, by DOE through grant DE-FG02-91ER40681 and by the SLOAN Foundation.  

\section{Appendix}

In this appendix we derive a useful identity for the theta functions. Instead of doing a general derivation we show how it works in the example
we are dealing with in the main text and then the generalization should be clear. 
Consider the function 
\beq
h(\lambda) = c \left(\frac{e^{i\pi\Delta_1^t \int_{p_1}^\lambda} \theta(a+\int_{p_1}^\lambda)}{\theta(\int_{p_1}^\lambda)}\right)^2\ ,
\eeq
defined on the Riemann surface. Also $c$ is a constant that we choose later. By looking at the results (\ref{Ch}) one can see that the numerator has zeros at $\lambda=0, \half(1-i), -1+i$ whereas
the denominator vanishes for $\lambda=\infty,\half(1-i), -1+i$. It follows that $h(\lambda)$ has a zero at $\lambda=0$ and a pole at $\lambda=\infty$. 
To see the behavior near $\lambda=0$ we use that (remembering that $p_1=0$ and eq.(\ref{wdef}))
\beq
\int_0^{\lambda} \omega_j \simeq  \int \frac{d\lambda}{i\sqrt{\lambda}} \lambda C^{-1}_{1j} = -2 i \sqrt{\lambda} C^{-1}_{1j}, \ \ \ \lambda\rightarrow 0 .
\eeq
Thus,
\beq
 h(\lambda) \simeq  - 4 c \lambda \left(\frac{D_{p_1}\theta(a)}{\theta(0)}\right)^2 .
\eeq
If we choose
\beq
 c = -\frac{1}{4} \left(\frac{\theta(0)}{D_{p_1}\theta(a)}\right)^2\ ,
 \eeq
 then $h(\lambda)\simeq \lambda$, that is, it has a simple zero at $\lambda=0$. Similarly one can check that it has a simple pole at $\lambda=\infty$. Finally it
 can be seen to have the right periodicity properties to be well defined on the Riemann surface. The only such function is $h(\lambda)=\lambda$ so we conclude
\beq
  \lambda = -\frac{1}{4} \left(\frac{\theta(0)}{D_{p_1}\theta(a)}\right)^2  
   \left(\frac{e^{i\pi\Delta_1^t \int_{p_1}^\lambda} \theta(a+\int_{p_1}^\lambda)}{\theta(\int_{p_1}^\lambda)}\right)^2\ ,
 \eeq
 as used in the main text. Furthermore, by expanding around $\lambda=\infty$ we find that $h(\lambda)=\lambda$ if we choose the constant $c$ to be
\beq
c =  -4 e^{i\pi\Delta_1^t\Pi\Delta_1} \left(\frac{D_{p_3}\theta(a)}{\theta(0)}\right)^2 .
\eeq
Equating the two expressions for $c$ we find
\beq
\left(\frac{D_{p_3}\theta(a)D_{p_1}\theta(a)}{\theta^2(0)}\right)^2 = \frac{1}{16} e^{-i\pi\Delta_1^t\Pi\Delta_1}\ ,
\eeq
as we also used in the main text. Although we derived the result for our particular case it is clearly valid in general (see for example \cite{FK} for the case
of a generic hyperelliptic Riemann surface). Taking the square root of the last equation we find
\beq
\frac{D_{p_3}\theta(a)D_{p_1}\theta(a)}{\theta^2(0)} = \pm \frac{1}{2} e^{-\frac{i}{2}\pi\Delta_1^t\Pi\Delta_1} .\label{idch}
\eeq
The sign cannot be determined by this reasoning but it is easily found to be minus by a simple numerical computation. The result was used in the main
text to find the correct normalization of $\zeta$ so that $\alpha$ is a solution to the cosh-Gordon equation.

\end{document}